\documentclass[onecolumn,showpacs,floatfix, aps, nofootinbib]{revtex4}
\usepackage{amssymb,amsmath}
\usepackage[dvips]{graphics,color}
\usepackage{epsfig}\usepackage{float}
\newcommand{\ba}{\begin{eqnarray}}
\newcommand{\ea}{\end{eqnarray}}
\newcommand{\bege}{\begin{equation}}
\newcommand{\bpartial}{\mathop{\partial\kern -4pt\raisebox{.8pt}{$|$}}}
\newcommand{\enge}{\end{equation}}
\newcommand{\beq}{\begin{eqnarray}}
\newcommand{\benu}{\begin{enumerate}}
\newcommand{\slg}{\mbox{\bfseries\slshape g}}
\newcommand{\be}{\begin{equation}}
\newcommand{\ee}{\end{equation}}
\newcommand{\enu}{\end{enumerate}}
\newcommand{\eeq}{\end{eqnarray}}
\newcommand{\pa}{\partial}

\newcommand{\RR}{\mathbb{R}}

\newcommand{\mk}{\mathfrak}

\newcommand{\me}{\frac{1}{2}}
\begin{document}

\title{Possible Generalizations within Braneworld Scenarios: Torsion fields}

\author{J. M. Hoff da Silva}
\email{hoff@feg.unesp.br} \affiliation{UNESP - Campus de Guaratinguet\'a - DFQ \\Av. Dr.
Ariberto Pereira da Cunha, 333\\ CEP 12516-410, Guaratinguet\'a - SP,
Brazil}
\author{Rold\~ao da Rocha}
\email{roldao.rocha@ufabc.edu.br}
\affiliation{ Centro de Matem\'atica, Computa\c c\~ao e Cogni\c
c\~ao,
Universidade Federal do ABC, 09210-170, Santo Andr\'e, SP, Brazil}


\begin{abstract}

In this Chapter we introduce the aspects in which torsion can influence the formalism of braneworld scenarios
in general, and also how it is possible to measure such kind of effects, namely, for instance, the blackstring transverse area corrections and variation of quasar luminosity due to those corrections.
We analyze the projected effective Einstein equation in a
$4$-dimensional arbitrary manifold embedded in a $5$-dimensional
Riemann-Cartan manifold. The Israel-Darmois matching conditions
are investigated, in the context where the torsion discontinuity
is orthogonal to the brane. Unexpectedly, the presence of torsion
terms in the connection does not modify such conditions
whatsoever, despite of the modification in the extrinsic curvature
and in the connection. Then, by imposing the
$\mathbb{Z}_{2}$-symmetry, the Einstein equation obtained via
Gauss-Codazzi formalism is extended, in order to now encompass the torsion terms.
We also show that the factors involving contorsion change
drastically the effective Einstein equation on the brane, as well
as the effective cosmological constant.
Also, we present gravitational aspects of braneworld models endowed with
torsion terms both in the bulk and on the brane. In order to
investigate a conceivable and measurable gravitational effect, arising genuinely
from bulk torsion terms, we analyze the variation in the black hole
area by the presence of torsion. Furthermore, we extend the
well known results about consistency conditions in a framework that incorporates brane
torsion terms. It is shown, in a rough estimate, that the resulting effects
are generally suppressed by the internal space volume.
This formalism provides manageable models and their possible ramifications into
some aspects of gravity in this context, and cognizable corrections and physical effects as well.
The torsion influences the braneworld scenario and we can check it by developing the bulk metric Taylor expansion around the brane, which brings corrections in the blackstring transverse area. This generalization is presented in order to better probe braneworld properties in a Riemann-Cartan framework, and it is also shown how the factors involving contorsion change the effective Einstein equation on the brane, the effective cosmological constant, and their consequence in a Taylor expansion of the bulk metric around the brane.

\end{abstract}
\maketitle
\section{Introduction}

In the last years there has been an increasing interest in large
extra dimensional models \cite{GVALI}, mainly due  the
developments in string theory \cite{HW} but also to the
possibility of the hierarchy problem explanation, presented for instance in Randall-Sundrum braneworld
scenarios \cite{RS}. In particular, the Randall-Sundrum braneworld
model \cite{RS} is effectively implemented in a 5-dimensional
manifold (where there is one warped extra dimension) and it is
based on a 5-dimensional reduction of Ho$\check{\mathrm
r}$ava-Witten theory. In Randall-Sundrum models,
our universe is described by an infinitely thin membrane
--- the brane. One attempt of explaining why gravity is so weak is
by trapping the braneworld in some higher-dimensional spacetime
--- the bulk --- wherein the brane is viewed as a submanifold. For
instance, the observable universe proposed by Randall and Sundrum,
in one of their two models, can be described as being a brane
embedded in an AdS$_5$ bulk.   There are several analogous models,
which consider our universe
as a $D$-dimensional braneworld embedded in a bulk of codimension 
 one.
In some models, there are some changes in the scenario that allows
the presence of a compact dimension on the brane \cite{PROP}. This gives rise
to the so called hybrid compactification.

As a
crucial formal pre-requisite to try to describe gravity in a
braneworld context, the bulk is imposed to present codimension one --- in relation to the brane.
There is a great amount of results applying
the Gauss-Codazzi (GC) formalism \cite{WALD} in order to derive
the properties of such braneworld (see \cite{JP} and references
therein). In the case where the bulk has two dimensions more than the brane, the GC
formalism is no longer useful, since the concept of a thin
membrane is meaningless, in the sense that it is not possible
to define junction conditions in codimension
greater than one. In such case the addition of a Gauss-Bonnet term
seems to break the braneworld apparent sterility \cite{RUTH}. For
higher codimensions, the situation is even worse.

Going back to the case of one non-compact extra dimension, after expressing the Einstein tensor in
terms of the stress tensor of the bulk and extrinsic curvature
corrections, it is
necessary to develop some mechanism to explore some physical
quantities on the brane. In order that the GC formalism be useful, we must be
able to express the quantities in the limit of the extra dimension
going to zero --- at the point where the brane is located. Using
this procedure,  two junction requirements
\cite{ISRAEL}, which are the well known Israel-Darmois matching conditions, emerge.
Moreover, torsion also emerges in the interface between GR and
gravity obtained via string theory at low energy. In this vein, it seems
quite natural to explore some aspects of braneworld models in the
presence of torsion. This is one of the main purposes of this Chapter,
where the matching conditions are analyzed and investigated
in the context of a braneworld of codimension one, described by a Riemann-Cartan
manifold, encoding torsion terms.

This Chapter is organized as follows: after presenting some
geometric preliminaries involving Riemann-Cartan spacetimes in the Section II, in Section III  the concept of torsion is introduced
in the context of general
relativity and the Israel-Darmois matching
conditions are investigated in the presence of torsion, in a similar approach that
can be found in reference \cite{PM}. In addition, junction
conditions are investigated in the context where the torsion
discontinuity is {orthogonal} to the brane.  In
Section IV the Gauss-Codazzi formalism is used in order to
establish the role and implications of torsion terms in the
braneworld framework scenario. All the quantities, like the Riemann and Ricci tensors, and the scalar curvature, and also the Einstein tensor taking into account torsion terms
are written in terms of their respective partners defined in terms of the Levi-Civita connection.

In order to find some typical gravitational signatures of
braneworld scenarios with torsion we obtain all the formul\ae\,
for a Taylor expansion outside a black hole in Section V,
extending some results of Ref. \cite{Maartens} in order to
encompass accrued torsion corrections. It is shown how the
contortion and its derivatives determine the variation in the area
of the black hole horizon along the extra dimension, inducing
observable physical effects.
 The Taylor expansion outside a black hole metric gives information about the bulk torsion terms, where the
corrections in the area of the 5$D$ black string horizon are
evinced.

In  Section VI we apply the braneworld consistency conditions in
the case when torsion is present in the brane manifold. We are
particularly concerned with the viability of such an extension,
analyzing the torsion effects in the brane scalar curvature. We
show that, for factorizable metrics, the torsion contribution to
the brane curvature is damped by the distance between the branes.
In warped braneworld models, however, this damping is --- at least
partially --- compensated by terms of the warp factor. It is also
shown that if the brane manifold is endowed with a connection
presenting torsion then a Randall-Sundrum like scenario with equal
sign brane tension becomes possible, in acute contrast to the
standard Randall-Sundrum model. Roughly speaking, this last
possibility comes from the following reasoning. The presence of
the torsion terms generally relax the consistence conditions,
specially in what concerns the sum over the brane tensions.
Therefore, the brane tensions are not restricted to the same sign,
although constrained by a specific contraction of contortion
terms. By studying the general consistency
conditions applied to this case, we arrive at some roughly
estimates concerning brane torsion effects. Although the presence
of torsion is not prohibited at all, its effects are generally
suppressed.

\section{Preliminaries}

\subsection{Classification of Metric Compatible Structures $(M,%
\slg
,D)$}

Let $M$ denote a $n$-dimensional manifold\footnote{We left the topology of $M$
unspecified for while.}. We denote as usual by $T_{x}M$ and $T_{x}^{\ast}M$
\ respectively the tangent and the cotangent spaces at $x\in M$. By $TM=
{\displaystyle\bigcup\nolimits_{x\in M}}
$ $T_{x}M$ and $T^{\ast}M=
{\displaystyle\bigcup\nolimits_{x\in M}}
$ $T_{x}^{\ast}M$ respectively the tangent and cotangent bundles. The spaces $T_{s}^{r}M$
we denote the bundle of $r$-contravariant and $s$-covariant tensors and by
$\mathcal{T}M=\bigoplus\nolimits_{r,s=0}^{\infty}T_{s}^{r}M$ the tensor
bundle. By $
{\displaystyle\bigwedge^{r}}
TM$ and $
{\displaystyle\bigwedge^{r}}
T^{\ast}M$ denote respectively the bundles of $r$-multivector fields and of
$r$-form fields. We call $
{\displaystyle\bigwedge}
TM=
{\displaystyle\bigoplus\nolimits_{r=0}^{r=n}}
{\displaystyle\bigwedge^{r}}
TM$ the bundle of (non homogeneous) multivector fields and call $%
{\displaystyle\bigwedge}
T^{\ast}M=
{\displaystyle\bigoplus\nolimits_{r=0}^{r=n}}
{\displaystyle\bigwedge^{r}}
T^{\ast}M$ the exterior algebra (Cartan) bundle. Of course, it is the bundle
of (non homogeneous) form fields. Recall that the real vector spaces are such
that $\dim\bigwedge^{r}T_{x}M$ $=\dim\bigwedge^{r}T_{x}^{\ast}M$ $=\binom
{n}{r}$ and $\dim
{\displaystyle\bigwedge}
T^{\ast}M=2^{n}$. Some \textit{additional} structures will be introduced or
mentioned below when needed. Let\footnote{We denote by$\ \sec(X(M))$ the space
of the sections of a bundle $X(M)$. Note that all functions and differential
forms are supposed smooth, unless we explicitly say the contrary.} $
\slg
\in\sec T_{2}^{0}M$ a metric of signature $(p,q)$ and $D$ an arbitrary metric
compatible connection on $M$, i.e., $D
\slg
=0$. We denote by $\mathbf{R}$ and $\mathbf{T}$ respectively the \ (Riemann)
curvature and torsion tensors\footnote{The precise definitions of those
objects will be recalled below.} of the connection $D$, and recall that in
general a given manifold given some additional conditions may admit many
different metrics and many different connections.

Given a triple $(M,
\slg, D)$,
\begin{enumerate}
\item[(a)] it is called a Riemann-Cartan space if and only if
$D\slg=0,\;{and}\;\mathbf{T}\neq 0,$
\item[(b)] it is called Weyl space if and only if
$D\slg\neq 0\;\mathrm{and}\;\mathbf{T}=0$,
\item[(c)] it is called a Riemann space if and only if
$D\slg=0\;\mathrm{and}\;\mathbf{T}=0, $
and in that case the pair $(D,\slg)$ is called \textit{Riemannian structure}.
\item[(d)] it is called a Riemann-Cartan-Weyl space if and only if
$D\slg\neq 0\;\mathrm{and}\;\mathbf{T}\neq 0$,
\item[(e)] it is called Riemann flat if and only if
$D\slg=0\;\mathrm{and}\;\mathbf{R}=0$,
\item[(f)] it is called teleparallel if and only if
$D\slg=0,\;\mathbf{T}\neq0\;\mathrm{and}\;\mathbf{R=}0$.
\end{enumerate}
\subsection{Levi-Civita and Riemann-Cartan Connections}
For each metric tensor defined on the manifold $M$ there exists one and only
one connection in the conditions of the item c) above. It is is called the
\textit{Levi-Civita connection\/} of the metric considered, and is denoted in
what follows by $\mathring{D}$. A connection satisfying the properties in (a)
above is called a Riemann-Cartan connection. In general both connections may
be defined in a given manifold and they are related by well established
formulas recalled below. \ A connection defines a rule for the parallel
transport of vectors (more generally tensor fields) in a manifold, something
which is conventional \cite{poincare}, and so the question concerning which
one is more important is according to our view meaningless.
\subsection{Spacetime Structures}

When $\dim M=4$ and the metric $\slg$ has signature $(1,3)$ we sometimes substitute Riemann by Lorentz in the
previous definitions \emph{(c)},\emph{(e)} and \emph{(f)}.
In order to represent a spacetime structure a Lorentzian or a Riemann-Cartan
structure $(M,%
\slg
,D)$ need be such that $M$ is connected and paracompact
and equipped with an orientation defined by the volume element $\tau_{%
\slg
}\in\sec%
{\displaystyle\bigwedge\nolimits^{4}}
T^{\ast}M$ and a time orientation denoted by $\uparrow$. We omit here the
details and ask to the interested reader to consult, e.g.,
\emph{\cite{rodoliv2007}}. A general spacetime will be represented by a
pentuple $(M,%
\slg
,D,\tau_{%
\slg
},\uparrow).$
{}

We call in what follows Hodge bundle the quadruple $(%
{\displaystyle\bigwedge}
T^{\ast}M,\wedge,\cdot,\tau_{%
\slg
})$ and now recall the meaning of the above symbols.

We suppose in what follows that any reader of this paper knows the meaning of
the exterior product of form fields and its main properties\footnote{We use
the conventions of \cite{rodoliv2007}.}. We simply recall\ here that if
$\mathcal{A}_{r}\in\sec\bigwedge^{r}T^{\ast}M$, $\mathcal{B}_{s}\in
\sec\bigwedge^{s}T^{\ast}M$ then
\begin{equation}
\mathcal{A}_{r}\wedge\mathcal{B}_{s}=(-1)^{rs}\mathcal{B}_{s}\wedge
\mathcal{A}_{r}. \label{exterior}%
\end{equation}
Let be $\mathcal{A}_{r}=a_{1}\wedge...\wedge a_{r}$ $\in\sec\bigwedge
^{r}T^{\ast}M$, $\mathcal{B}_{r}=b_{1}\wedge...\wedge b_{r}\in\sec
\bigwedge^{r}T^{\ast}M$ \ where $a_{i},b_{j}\in\sec\bigwedge^{1}T^{\ast}M$
$(i,j=1,2,...,r).$

(i) The scalar product $\mathcal{A}_{r}\cdot\mathcal{B}_{r}$ is defined by
\begin{align}
\mathcal{A}_{r}\cdot\mathcal{B}_{r}  &  =(a_{1}\wedge...\wedge a_{r}%
)\cdot(b_{1}\wedge...\wedge b_{r})\nonumber\\
&  =\left\vert
\begin{array}
[c]{lll}%
a_{1}\cdot b_{1} & .... & a_{1}\cdot b_{r}\\
.......... & .... & ..........\\
a_{r}\cdot b_{1} & .... & a_{r}\cdot b_{r}%
\end{array}
\right\vert . \label{scalarprod}%
\end{align}
where $a_{i}\cdot b_{j}:=\mathtt{g}(a_{i},b_{j})$.

We agree that if $r=s=0$, the scalar product is simple the ordinary product in
the real field.

Also, if $r\neq s$, then $\mathcal{A}_{r}\cdot\mathcal{B}_{s}=0$. Finally, the
scalar product is extended by linearity for all sections of $\mathcal{%
{\displaystyle\bigwedge}
}T^{\ast}M$.

For $r\leq s,$ $\mathcal{A}_{r}=a_{1}\wedge...\wedge a_{r},$ $\mathcal{B}%
_{s}=b_{1}\wedge...\wedge b_{s\text{ }}$we define the \textit{left
contraction} by
\begin{equation}
\lrcorner:(\mathcal{A}_{r},\mathcal{B}_{s})\mapsto\mathcal{A}_{r}%
\lrcorner\mathcal{B}_{s}=%
{\displaystyle\sum\limits_{i_{1}\,<...\,<i_{r}}}
\epsilon^{i_{1}....i_{s}}(a_{1}\wedge...\wedge a_{r})\cdot(b_{_{i_{1}}}%
\wedge...\wedge b_{i_{r}})^{\sim}b_{i_{r}+1}\wedge...\wedge b_{i_{s}}
\end{equation}
where $\sim$ is the reverse mapping (\emph{reversion}) defined by
\begin{equation}
\sim:\sec%
{\displaystyle\bigwedge\nolimits^{p}}
T^{\ast}M\ni a_{1}\wedge...\wedge a_{p}\mapsto a_{p}\wedge...\wedge a_{1}
\end{equation}
and extended by linearity to all sections of $%
{\displaystyle\bigwedge}
T^{\ast}M$. We agree that for $\alpha,\beta\in\sec\bigwedge^{0}T^{\ast}M$ the
contraction is the ordinary (pointwise) product in the real field and that if
$\alpha\in\sec\bigwedge^{0}T^{\ast}M$, $\mathcal{A}_{r}\in\sec\bigwedge
^{r}T^{\ast}M$, $\mathcal{B}_{s}\in\sec\bigwedge^{s}T^{\ast}M$ then
$(\alpha\mathcal{A}_{r})\lrcorner\mathcal{B}_{s}=\mathcal{A}_{r}%
\lrcorner(\alpha\mathcal{B}_{s})$. Left contraction is extended by linearity
to all pairs of elements of sections of $%
{\displaystyle\bigwedge}
T^{\ast}M$, i.e., for $\mathcal{A},\mathcal{B}\in\sec%
{\displaystyle\bigwedge}
T^{\ast}M$%

\begin{equation}
\mathcal{A\lrcorner B}=\sum_{r,s}\langle\mathcal{A}\rangle_{r}\lrcorner
\langle\mathcal{B}\rangle_{s},\text{ }r\leq s, \label{9bis}%
\end{equation}
where $\langle\mathcal{A}\rangle_{r}$ means the projection of $\mathcal{A}$ in
$%
{\displaystyle\bigwedge\nolimits^{r}}
T^{\ast}M$.

It is also necessary to introduce the operator of \emph{right contraction}
denoted by $\llcorner$. The definition is obtained from the one presenting the
left contraction with the imposition that $r\geq s$ and taking into account
that now if $\mathcal{A}_{r}\in\sec\bigwedge^{r}T^{\ast}M$, $\mathcal{B}%
_{s}\in\sec\bigwedge^{s}T^{\ast}M\ $then $\mathcal{B}_{s}\lrcorner
\mathcal{A}_{r}=(-1)^{s(r-s)}\mathcal{A}_{r}\llcorner\mathcal{B}_{s}$.

\subsection{Exterior derivative $d$ and Hodge coderivative $\delta$}

The \textit{exterior derivative is a mapping\/}
\[
d:\sec\bigwedge T^{\ast}M\rightarrow\sec\bigwedge T^{\ast}M,
\]
satisfying:
\begin{equation}%
\begin{array}
[c]{ll}%
\text{(i)} & d(A+B)=dA+dB;\\
\text{(ii)} & d(A\wedge B)=dA\wedge B+\bar{A}\wedge dB;\\
\text{(iii)} & df(v)=v(f);\\
\text{(iv)} & d^{2}=0,
\end{array}
\label{ext deriv}%
\end{equation}
for every $A,B\in\sec\bigwedge T^{\ast}M$, $f\in\sec\bigwedge^{0}T^{\ast}M$
and $v\in\sec TM$.

The \textit{Hodge codifferential} operator in the Hodge bundle is the mapping
$\delta:\sec\bigwedge^{r}T^{\ast}M\rightarrow\sec\bigwedge^{r-1}T^{\ast}M$,
given for homogeneous multiforms, by:%
\begin{equation}
\delta=(-1)^{r}\star^{-1}d\star, \label{hodge}%
\end{equation}
where $\star$ is the Hodge star operator. The operator $\delta$ extends by
linearity to all $\bigwedge T^{\ast}M$.
\subsection{Clifford Bundles}

Let $(M,%
\slg
,\nabla)$ be a Riemannian, Lorentzian or Riemann-Cartan
structure\footnote{$\nabla$ may be the Levi-Civita connection $\mathring{D}$
of $%
\slg
$ or an arbitrary Riemann-Cartan connection $D$.}. As before let
$\mathtt{g}\in\sec T_{0}^{2}M$ be the metric on the cotangent bundle
associated with $%
\slg
\in\sec T_{2}^{0}M$. Then $T_{x}^{\ast}M\simeq\mathbb{R}^{p,q}$, where
$\mathbb{R}^{p,q}$ is a vector space equipped with a scalar product
$\bullet\equiv\left.  \mathtt{g}\right\vert _{x}$ of signature $(p,q)$. The
Clifford bundle of differential forms $\mathcal{C}\!\ell(M,\mathtt{g})$ is the
bundle of algebras, i.e., $\mathcal{C}\ell(M,\mathtt{g})=\cup_{x\in
M}\mathcal{C}\ell(T_{x}^{\ast}M,\bullet)$, where $\forall x\in M$,
$\mathcal{C}\!\ell(T_{x}^{\ast}M,\mathbf{\bullet})=\mathbb{R}_{p,q}$, a real
Clifford algebra. When the structure $(M,%
\slg
,\nabla)$ is part of a Lorentzian or Riemann-Cartan spacetime $\mathcal{C}%
\ell(T_{x}^{\ast}M,\mathbf{\bullet})=\mathbb{R}_{1,3}$ the so called
\emph{spacetime} \emph{algebra}. Recall also that $\mathcal{C}\ell
(M,\mathtt{g})$ is a vector bundle associated with the \emph{\ }$\mathtt{g}%
$-\emph{orthonormal coframe bundle \ }$\mathbf{P}_{\mathrm{SO}_{(p,q)}^{e}%
}(M,\mathtt{g})$, i.e., $\mathcal{C}\ell(M,\mathtt{g})$ $=P_{\mathrm{SO}%
_{(p,q)}^{e}}(M,\mathtt{g})\times_{ad}\mathbb{R}_{1,3}$ (see more details in,
e.g., \cite{mosnawal,rodoliv2007}). For any $x\in M$, $\mathcal{C}\ell
(T_{x}^{\ast}M,\bullet)$ is a linear space over the real field $\mathbb{R}$.
Moreover, $\mathcal{C}\ell(T_{x}^{\ast}M)$ is isomorphic as a real vector
space to the Cartan algebra $\bigwedge T_{x}^{\ast}M$ of the cotangent space.
Then, sections of $\mathcal{C}\ell(M,\mathtt{g})$ can be represented as a
\textit{sum} of non homogeneous differential forms. Let now $\{\mathbf{e}%
_{\mathbf{a}}\}$ be an orthonormal basis for $TU$ and $\{\theta^{\mathbf{a}%
}\}$ its dual basis. Then, \texttt{g}$(\theta^{\mathbf{a}},\theta^{\mathbf{b}%
})=\eta^{\mathbf{ab}}.$

The fundamental \emph{Clifford product} (in what follows to be denoted by
juxtaposition of symbols) is generated by%
\begin{equation}
\theta^{\mathbf{a}}\theta^{\mathbf{b}}+\theta^{\mathbf{b}}\theta^{\mathbf{a}%
}=2\eta^{\mathbf{ab}} \label{cp}%
\end{equation}
and if $\mathcal{C}\in\mathcal{C}\ell(M,\mathtt{g})$ we have%

\begin{equation}
\mathcal{C}=s+v_{\mathbf{a}}\theta^{\mathbf{a}}+\frac{1}{2!}b_{\mathbf{ab}%
}\theta^{\mathbf{a}}\theta^{\mathbf{b}}+\frac{1}{3!}a_{\mathbf{abc}}%
\theta^{\mathbf{a}}\theta^{\mathbf{b}}\theta^{\mathbf{c}}+p\theta^{n+1}\;,
\end{equation}
where $\tau_{%
\slg
}:=\theta^{n+1}=\theta^{0}\theta^{1}\theta^{2}\theta^{3}...\theta^{n}$ is the
volume element and $s$, $v_{\mathbf{a}}$, $b_{\mathbf{ab}}$, $a_{\mathbf{abc}%
}$, $p\in\sec\bigwedge^{0}T^{\ast}M\hookrightarrow\sec\mathcal{C}%
\ell(M,\mathtt{g})$.

Let $\mathcal{A}_{r},\in\sec\bigwedge^{r}T^{\ast}M\hookrightarrow
\sec\mathcal{C}\!\ell(M,\mathtt{g}),\mathcal{B}_{s}\in\sec\bigwedge^{s}%
T^{\ast}M\hookrightarrow\sec\mathcal{C}\ell(M,\mathtt{g})$. For $r=s=1$, we
define the \emph{scalar product} as follows:

For $a,b\in\sec\bigwedge^{1}T^{\ast}M\hookrightarrow\sec\mathcal{C}%
\ell(M,\mathtt{g}),$%
\begin{equation}
a\cdot b=\frac{1}{2}(ab+ba)=\mathtt{g}(a,b).
\end{equation}
We identify the \emph{exterior product} ($\forall r,s=0,1,2,3)$ of homogeneous
forms (already introduced above) by
\begin{equation}
\mathcal{A}_{r}\wedge\mathcal{B}_{s}=\langle\mathcal{A}_{r}\mathcal{B}%
_{s}\rangle_{r+s},
\end{equation}
where $\langle\rangle_{k}$ is the \textit{component} in $\bigwedge^{k}T^{\ast
}M$ \ (projection) of the Clifford field. The exterior product is extended by
linearity to all sections of $\mathcal{C}\ell(M,\mathtt{g})$.

The scalar product, the left and the right are defined for homogeneous form
fields that are sections of the Clifford bundle in exactly the same way as in
the Hodge bundle and they are extended by linearity for all sections of
$\mathcal{C}\ell(M,\mathtt{g})$.

In particular, for $\mathcal{A},\mathcal{B}\in\sec\mathcal{C}\ell
(M,\mathtt{g})$ we have%

\begin{equation}
\mathcal{A\lrcorner B}=\sum_{r,s}\langle\mathcal{A}\rangle_{r}\lrcorner
\langle\mathcal{B}\rangle_{s},\text{ }r\leq s.
\end{equation}

The main formulas used in the present paper can be obtained (details may be
found in \cite{rodoliv2007}) from the following ones (where $a\in\sec
\bigwedge^{1}T^{\ast}M\hookrightarrow\sec\mathcal{C}\!\ell(M,\mathtt{g})$):
\begin{align}
a\mathcal{B}_{s}  &  =a\lrcorner\mathcal{B}_{s}+a\wedge\mathcal{B}%
_{s},\;\;\mathcal{B}_{s}a=\mathcal{B}_{s}\llcorner a+\mathcal{B}_{s}\wedge
a,\nonumber\\
a\lrcorner\mathcal{B}_{s}  &  =\frac{1}{2}(a\mathcal{B}_{s}-(-1)^{s}%
\mathcal{B}_{s}a),\nonumber\\
\mathcal{A}_{r}\lrcorner\mathcal{B}_{s}  &  =(-1)^{r(s-r)}\mathcal{B}%
_{s}\llcorner\mathcal{A}_{r},\nonumber\\
a\wedge\mathcal{B}_{s}  &  =\frac{1}{2}(a\mathcal{B}_{s}+(-1)^{s}%
\mathcal{B}_{s}a),\nonumber\\
\mathcal{A}_{r}\mathcal{B}_{s}  &  =\langle\mathcal{A}_{r}\mathcal{B}%
_{s}\rangle_{|r-s|}+\langle\mathcal{A}_{r}\mathcal{B}_{s}\rangle
_{|r-s|+2}+...+\langle\mathcal{A}_{r}\mathcal{B}_{s}\rangle_{|r+s|}\nonumber\\
&  =\sum\limits_{k=0}^{m}\langle\mathcal{A}_{r}\mathcal{B}_{s}\rangle
_{|r-s|+2k}\text{ }\nonumber\\
\mathcal{A}_{r}\cdot\mathcal{B}_{r}  &  =\mathcal{B}_{r}\cdot\mathcal{A}%
_{r}=\widetilde{\mathcal{A}}_{r}\text{ }\lrcorner\mathcal{B}_{r}%
=\mathcal{A}_{r}\llcorner\widetilde{\mathcal{B}}_{r}=\langle\widetilde
{\mathcal{A}}_{r}\mathcal{B}_{r}\rangle_{0}=\langle\mathcal{A}_{r}%
\widetilde{\mathcal{B}}_{r}\rangle_{0},\nonumber\\
\star\mathcal{A}_{k}  &  =\widetilde{\mathcal{A}}_{k}\lrcorner\tau_{%
\slg
}=\widetilde{\mathcal{A}}_{k}\tau_{%
\slg
}. \label{10}%
\end{align}
Two other important identities to be used below are:%

\begin{equation}
a\lrcorner(\mathcal{X}\wedge\mathcal{Y})=(a\lrcorner\mathcal{X})\wedge
\mathcal{Y}+\mathcal{\hat{X}}\wedge(a\lrcorner\mathcal{Y}), \label{T54}%
\end{equation}
for any $a\in\sec%
{\displaystyle\bigwedge\nolimits^{1}}
T^{\ast}M$ and $\mathcal{X},\mathcal{Y}\in\sec%
{\displaystyle\bigwedge}
T^{\ast}M$, and
\begin{equation}
A\lrcorner(B\lrcorner C)=(A\wedge B)\lrcorner C, \label{T50}%
\end{equation}
for any $A,B,C\in\sec\bigwedge T^{\ast}M\hookrightarrow\mathcal{C}%
\ell(M,\mathtt{g})$.

\subsection{Torsion, Curvature and Cartan Structure Equations}

As we said in the beginning of Section 1 a given structure $(M,%
\slg
)$ may admit many different metric compatible connections. Let then
$\mathring{D}$ be the Levi-Civita connection of $%
\slg
$ and $D$ a Riemann-Cartan connection acting on the tensor fields defined on
$M$.

Let $U\subset M$ and consider a chart of the maximal atlas of $M$ covering $U$
with arbitrary coordinates $\{x^{\mu}\}$. Let $\{{\mbox{\boldmath$\partial$}}%
_{\mu}\}$ be a basis for $TU$, $U\subset M$ and let $\{\theta^{\mu}=dx^{\mu
}\}$ be the dual basis of $\{{\mbox{\boldmath$\partial$}}_{\mu}\}$. The
reciprocal basis of $\{\theta^{\mu}\}$ is denoted $\{\theta^{\mu}\}$, and
\texttt{g}$(\theta^{\mu},\theta_{\nu}):=$ $\theta^{\mu}\cdot\theta_{\nu
}=\delta_{\nu}^{\mu}$.

Let also $\{\mathbf{e}_{\mathbf{a}}\}$ be an orthonormal basis for $TU\subset
TM$ \ with $\mathbf{e}_{\mathbf{b}}=q_{\mathbf{b}}^{\nu}%
{\mbox{\boldmath$\partial$}}_{\nu}$.$\mathtt{\ }$The dual basis of $TU$ is
$\{\theta^{\mathbf{a}}\}$, with $\theta^{\mathbf{a}}=q_{\mu}^{\mathbf{a}%
}dx^{\mu}$. Also, $\{\theta_{\mathbf{b}}\}$ is the reciprocal basis of
$\{\theta^{\mathbf{a}}\}$, i.e. $\theta^{\mathbf{a}}\cdot\theta_{\mathbf{b}%
}=\delta_{\mathbf{b}}^{\mathbf{a}}$. An arbitrary frame on $TU\subset TM$,
coordinate or orthonormal will be denote by $\{e_{\alpha}\}$. Its dual frame
will be denoted by $\{\vartheta^{\rho}\}$ (i.e., $\vartheta^{\rho}(e_{\alpha
})=\delta_{\alpha}^{\rho}$ ).

\subsection{Torsion and Curvature Operators}

{}
The \textit{torsion and curvature operators }$\mathbf{\tau}$ and
$\mathbf{\rho}$ of a connection $D$, are respectively the mappings:%
\begin{align}
\mathbf{\tau}(\mathbf{u,v}) &  =D_{\mathbf{u}}\mathbf{v}-D_{\mathbf{v}%
}\mathbf{u}-[\mathbf{u,v}],\label{top}\\
\mathbf{\rho(u,v)} &  =D_{\mathbf{u}}D_{\mathbf{v}}-D_{\mathbf{v}%
}D_{\mathbf{u}}-D_{[\mathbf{u,v}]},\label{cop}%
\end{align}
for every $\mathbf{u,v}\in\sec TM$.
{}

\subsection{Torsion and Curvature Tensors}

{}
The \textit{torsion and\/curvature }tensors of a connection $D$, are
respectively the mappings:%

\begin{align}
\mathbf{T}(\alpha,\mathbf{u,v})  &  =\alpha\left(  \mathbf{\tau}%
(\mathbf{u},\mathbf{v})\right)  ,\label{to op}\\
\mathbf{R}(\mathbf{w},\alpha,\mathbf{u,v})  &  =\alpha(\mathbf{\rho(u,v)w}),
\label{curv op}%
\end{align}
for every $\mathbf{u,v,w}\in\sec TM$ and $\alpha\in\sec\bigwedge^{1}T^{\ast}M$.
{}

We recall that for any differentiable functions $f,g$ and $h$ we have%
\begin{align}
\mathbf{\tau}(g\mathbf{u,}h\mathbf{v})  &  =gh\mathbf{\tau}(\mathbf{u,v}%
),\nonumber\\
\mathbf{\rho(}g\mathbf{u,}h\mathbf{v)}f\mathbf{w}  &  \mathbf{=}%
ghf\mathbf{\rho(u,v)w} \label{exerctocur}%
\end{align}

\subsubsection{Properties of the Riemann Tensor for a Metric Compatible
Connection}

Note that it is quite obvious that
\begin{equation}
\mathbf{R}(\mathbf{w},\alpha,\mathbf{u,v})=\mathbf{R}(\mathbf{w}%
,\alpha,\mathbf{v,u}).
\end{equation}
Define the tensor field $\mathbf{R}^{\prime}$ as the mapping such that for
every $\mathbf{a,u,v,w}\in\sec TM$ and $\alpha\in\sec\bigwedge^{1}T^{\ast}M$.
\begin{equation}
\mathbf{R}^{\prime}(\mathbf{w},\mathbf{a},\mathbf{u,v})=\mathbf{R}%
(\mathbf{w},\alpha,\mathbf{v,u}). \label{ri3}%
\end{equation}
It is quite obvious that
\begin{equation}
\mathbf{R}^{\prime}(\mathbf{w},\mathbf{a},\mathbf{u,v})=\mathbf{a\cdot
}(\mathbf{\rho(u,v)w}), \label{ri4}%
\end{equation}
where
\begin{equation}
\alpha=%
\slg
(\mathbf{a,\;)}\text{, }\mathbf{a}=\mathtt{g}(\alpha,\;) \label{ri5}%
\end{equation}
We now show that for any structure $(M,%
\slg
,D)$ such that $D%
\slg
=0$ we have for $\mathbf{c},\mathbf{u,v}\in\sec TM$,
\begin{equation}
\mathbf{R}^{\prime}(\mathbf{c},\mathbf{c},\mathbf{u,v})=\mathbf{c\cdot
}(\mathbf{\rho(u,v)c})=0. \label{ri6}%
\end{equation}

We start recalling that for every metric compatible connection it holds:
\begin{align}
\mathbf{u(v(c\cdot c)} &  \mathbf{=u(}D_{\mathbf{v}}\mathbf{c\cdot c+c\cdot
}D_{\mathbf{v}}\mathbf{c)=}2\mathbf{u(}D_{\mathbf{v}}\mathbf{c\cdot
c)}\nonumber\\
&  =2\mathbf{(}D_{\mathbf{u}}D_{\mathbf{v}}\mathbf{c)\cdot c}+2\mathbf{(}%
D_{\mathbf{u}}\mathbf{c)\cdot D_{\mathbf{v}}c,}\label{ri7}%
\end{align}
Exchanging $\mathbf{u}\leftrightarrow\mathbf{v}$ in the last equation we get%
\begin{equation}
\mathbf{v(u(c\cdot c)=}2\mathbf{(}D_{\mathbf{v}}D_{\mathbf{u}}\mathbf{c)\cdot
c}+2\mathbf{(}D_{\mathbf{v}}\mathbf{c)\cdot D_{\mathbf{u}}c.}\label{ri8}%
\end{equation}
Subtracting Eq.(\ref{ri7}) from Eq.(\ref{ri8}) we have
\begin{equation}
\lbrack\mathbf{u,v](c\cdot c)=}2([D_{\mathbf{u}},D_{\mathbf{v}}%
]\mathbf{c)\cdot c}\label{ri9}%
\end{equation}
But since
\begin{equation}
\lbrack\mathbf{u,v](c\cdot c)=}D_{[\mathbf{u,v]}}(\mathbf{c\cdot
c})=2(D_{[\mathbf{u,v]}}\mathbf{c)\cdot c,}\label{ri10}%
\end{equation}
we have from Eq.(\ref{ri9}) that
\begin{equation}
([D_{\mathbf{u}},D_{\mathbf{v}}]\mathbf{c-}D_{[\mathbf{u,v]}}\mathbf{c)\cdot
c}=0\text{,}\label{ri11}%
\end{equation}
and it follows that $\mathbf{R}^{\prime}(\mathbf{c},\mathbf{c},\mathbf{u,v}%
)=0$ as we wanted to show.

{}
Prove that for any metric compatible connection,
\begin{equation}
\mathbf{R}^{\prime}(\mathbf{c},\mathbf{d},\mathbf{u,v})=\mathbf{R}^{\prime
}(\mathbf{d},\mathbf{c},\mathbf{u,v}). \label{ri12}%
\end{equation}

{}

Given an arbitrary frame $\{e_{\alpha}\}$ on $TU\subset TM$, let
$\{\vartheta^{\rho}\}$ be the \textit{dual frame}. We write:
\begin{equation}%
\begin{array}
[c]{ccl}%
\lbrack e_{\alpha}\mathbf{,}e_{\beta}] & = & c_{\alpha\beta}^{\rho}e_{\rho}\\
D_{e_{\alpha}}e_{\beta} & = & \mathbf{L}_{\alpha\beta}^{\rho}e_{\rho},
\end{array}
\end{equation}
where $c_{\alpha\beta}^{\rho}$ are the \textit{structure coefficients\/}of the
frame $\{e_{\alpha}\}$ and $\mathbf{L}_{\alpha\beta}^{\rho}$ are the
\textit{connection coefficients\/}in this frame. Then, the components of the
torsion and curvature tensors are given, respectively, by:
\begin{equation}%
\begin{array}
[c]{c}%
\mathbf{T}(\vartheta^{\rho},e_{\alpha}\mathbf{,}e_{\beta})=T_{\alpha\beta
}^{\rho}=\mathbf{L}_{\alpha\beta}^{\rho}-\mathbf{L}_{\beta\alpha}^{\rho
}-c_{\alpha\beta}^{\rho}\\
\mathbf{R(}e_{\mu},\vartheta^{\rho},e_{\alpha}\mathbf{,}e_{\beta}%
\mathbf{)}=R_{\mu}{}^{\rho}{}_{\!\alpha\beta}=e_{\alpha}(\mathbf{L}_{\beta\mu
}^{\rho})-e_{\beta}(\mathbf{L}_{\alpha\mu}^{\rho})+\mathbf{L}_{\alpha\sigma
}^{\rho}\mathbf{L}_{\beta\mu}^{\sigma}-\mathbf{L}_{\beta\sigma}^{\rho
}\mathbf{L}_{\alpha\mu}^{\sigma}-c_{\alpha\beta}^{\sigma}\mathbf{L}_{\sigma
\mu}^{\rho}.
\end{array}
\label{585}%
\end{equation}

It is important for what follows to keep in mind the definition of the
(symmetric) Ricci tensor, here denoted $\mathbf{Ric}\in\sec T_{2}^{0}M$ and
which in an arbitrary basis is written as
\begin{equation}
\mathbf{Ric=}R\mathbf{_{\mu\nu}\vartheta^{\mu}\otimes\vartheta^{\nu}:=}R_{\mu
}{}^{\rho}{}_{\!\rho\nu}\vartheta^{\mu}\otimes\vartheta^{\nu}. \label{Ricci}%
\end{equation}
It is crucial here to take into account the \textit{place} where the
contractions in the Riemann tensor takes place according to our conventions.

We also have:
\begin{equation}%
\begin{array}
[c]{l}%
d\vartheta^{\rho}=-\frac{1}{2}c_{\alpha\beta}^{\rho}\vartheta^{\alpha}%
\wedge\vartheta^{\beta},\\
D_{e_{\alpha}}\vartheta^{\rho}=-\mathbf{L}_{\alpha\beta}^{\rho}\vartheta
^{\beta},%
\end{array}
\label{608}%
\end{equation}
where $\omega_{\beta}^{\rho}\in\sec\bigwedge^{1}T^{\ast}M$ are the
\textit{connection 1-forms, }$\mathbf{L}_{\alpha\beta}^{\rho}$ \ are said to
be the connection coefficients in the given basis, and the $\mathcal{T}^{\rho
}\in\sec\bigwedge^{2}T^{\ast}M$ are the \textit{torsion 2-forms\/} and the
$\mathcal{R}_{\beta}^{\rho}\in\sec\bigwedge^{2}T^{\ast}M$\textbf{\ }are the
\textit{curvature 2-forms\/}, given by:
\begin{align}
&  \omega_{\beta}^{\rho}=\mathbf{L}_{\alpha\beta}^{\rho}\vartheta^{\alpha
},\nonumber\\
&  \mathcal{T}^{\rho}=\frac{1}{2}T_{\alpha\beta}^{\rho}\vartheta^{\alpha
}\wedge\theta^{\beta}\label{620}\\
&  \mathcal{R}_{\mu}^{\rho}=\frac{1}{2}R_{\mu}{}^{\rho}{}_{\!\alpha\beta
}\vartheta^{\alpha}\wedge\vartheta^{\beta}.\nonumber
\end{align}

Multiplying Eqs.(\ref{585}) by $\frac{1}{2}\vartheta^{\alpha}\wedge
\vartheta^{\beta}$ and using Eqs.(\ref{608}) and~(\ref{620}), we get the so-called  Cartan Structure Equations:

\begin{equation}%
\begin{array}
[c]{l}%
d\vartheta^{\rho}+\omega_{\beta}^{\rho}\wedge\vartheta^{\beta}=\mathcal{T}%
^{\rho},\\
d\omega_{\mu}^{\rho}+\omega_{\beta}^{\rho}\wedge\omega_{\mu}^{\beta
}=\mathcal{R}_{\mu}^{\rho}.
\end{array}
\label{559}%
\end{equation}

We can show that the torsion and (Riemann) curvature tensors can be written as%
\begin{align}
\mathbf{T}  &  =e_{\alpha}\otimes\mathcal{T}^{\alpha},\\
\mathbf{R}  &  =e_{\rho}\otimes e^{\mu}\otimes\mathcal{R}_{\mu}^{\rho}.
\end{align}

\subsection{Exterior Covariant Derivative $\mathbf{D}$}

Sometimes, Eqs.(\ref{559}) are written by some authors \cite{thiwal} as:
\begin{align}
\mathbf{D}\vartheta^{\rho} &  =\mathcal{T}^{\rho},\label{559a}\\
\text{\textquotedblleft\ }\mathbf{D}\omega_{\mu}^{\rho} &  =\mathcal{R}_{\mu
}^{\rho}.\text{\textquotedblright}\label{559b}%
\end{align}
and $\mathbf{D}:\sec\bigwedge T^{\ast}M\rightarrow\sec\bigwedge T^{\ast}M$ is
said to be the \textit{exterior covariant derivative}\/related to the
connection $D$. Now, Eq.(\ref{559b}) has been printed with quotation marks due
to the fact that it is an \textit{incorrect} equation. Indeed, a\textit{
legitimate} exterior covariant derivative operator\footnote{Sometimes also
called exterior covariant differential.} is a concept that can be defined for
$(p+q)$-indexed $r$-form fields\footnote{Which is not the case of the
connection $1$-forms $\omega_{\beta}^{\alpha}$, despite the name. More
precisely, the $\omega_{\beta}^{\alpha}$ are not true indexed forms, i.e.,
there does not exist a tensor field $\mathbf{\omega}$ such that
$\mathbf{\omega(}e_{i},e_{\beta},\vartheta^{\alpha})=$ $\omega_{\beta}%
^{\alpha}(e_{i}).$} as follows. Suppose that $X\in\sec T_{p}^{r+q}M$ and let
\begin{equation}
X_{\nu_{1}....\nu_{q}}^{\mu_{1}....\mu_{p}}\in\sec\bigwedge\nolimits^{r}%
T^{\ast}M,\label{559new}%
\end{equation}
such that for $v_{i}\in\sec TM,$ $i=0,1,2,..,r$,
\begin{equation}
X_{\nu_{1}....\nu_{q}}^{\mu_{1}....\mu_{p}}(v_{1},...,v_{r})=X(v_{1}%
,...,v_{r},e_{\nu_{1}},...,e_{\nu_{q}},\vartheta^{\mu_{1}},...,\vartheta
^{\mu_{p}}).\label{559new1}%
\end{equation}

The exterior covariant differential $\mathbf{D}$\textbf{ }of $X_{\nu
_{1}....\nu_{q}}^{\mu_{1}....\mu_{p}}$ on a manifold with a general connection
$D$ is the mapping:%

\begin{equation}
\mathbf{D:}\sec\bigwedge\nolimits^{r}T^{\ast}M\rightarrow\sec\bigwedge
\nolimits^{r+1}T^{\ast}M\text{, }0\leq r\leq4,\label{559new2}%
\end{equation}
such that\footnote{As usual the inverted hat over a symbol (in
Eq.(\ref{559new3})) means that the corresponding symbol is missing in the
expression.}%
\begin{align}
&  (r+1)\mathbf{D}X_{\nu_{1}....\nu_{q}}^{\mu_{1}....\mu_{p}}(v_{0}%
,v_{1},...,v_{r})\nonumber\\
&  =\sum\limits_{\nu=0}^{r}(-1)^{\nu}D_{\mathbf{e}_{\nu}}X(v_{0}%
,v_{1},...,\check{v}_{\nu},...v_{r},e_{\nu_{1}},...,e_{\nu_{q}},\vartheta
^{\mu_{1}},...,\vartheta^{\mu_{p}})\nonumber\\
&  -\sum\limits_{0\leq\lambda,\varsigma\,\leq r}(-1)^{\nu+\varsigma
}X(\mathbf{T(}v_{\lambda},v_{\varsigma}),v_{0},v_{1},...,\check{v}_{\lambda
},...,\check{v}_{\varsigma},...,v_{r},e_{\nu_{1}},...,e_{\nu_{q}}%
,\vartheta^{\mu_{1}},...,\vartheta^{\mu_{p}}).\label{559new3}%
\end{align}

Then, we may verify that
\begin{align}
\mathbf{D}X_{\nu_{1}....\nu_{q}}^{\mu_{1}....\mu_{p}}  &  =dX_{\nu_{1}%
....\nu_{q}}^{\mu_{1}....\mu_{p}}+\omega_{\mu_{s}}^{\mu_{1}}\wedge X_{\nu
_{1}....\nu_{q}}^{\mu_{s}....\mu_{p}}+...+\text{ }\omega_{\mu_{s}}^{\mu_{1}%
}\wedge X_{\nu_{1}....\nu_{q}}^{\mu_{1}....\mu_{p}}\label{559new4}\\
&  -\omega_{\nu_{1}}^{\nu_{s}}\wedge X_{\nu_{s}....\nu_{q}}^{\mu_{1}%
....\mu_{p}}-...-\text{ }\omega_{\mu_{s}}^{\mu_{1}}\wedge X_{\nu_{1}%
....\nu_{s}}^{\mu_{1}....\mu_{p}}.\nonumber
\end{align}

{}
Note that if \emph{Eq.(\ref{559new4})} is applied on any one of the connection
$1$-forms $\omega_{\nu}^{\mu}$ we would get $\mathbf{D}\omega_{\nu}^{\mu
}=d\omega_{\nu}^{\mu}+\omega_{\alpha}^{\mu}\wedge\omega_{\nu}^{\alpha}%
-\omega_{\nu}^{\alpha}\wedge\omega_{\alpha}^{\mu}$.

\section{Junction conditions}

In this Section some mathematical preliminaries --- necessary to
investigate braneworld junction conditions in a $D$-dimensional Riemann-Cartan
manifold, embedded in an arbitrary $(D+1)$-dimensional manifold ---  are briefly presented and discussed. For a
complete exposition concerning arbitrary manifolds and fiber
bundles, see, e.g, \cite{frankel, naka, koni, moro, rowa}.

Hereon $\Sigma$ denotes a $D$-dimensional Riemann-Cartan manifold
modeling a brane embedded in a bulk, denoted by $M$. A vector
space endowed with a constant signature  metric, isomorphic to
 $\RR^{D+1}$, can be identified at a point $x\in M$ as being the space
$T_xM$ tangent to $M$, where $M$ is locally diffeomorphic to its
own (local) foliation
 $\RR\times\Sigma$. There always exists a 1-form field $n$, normal to $\Sigma$, which can be locally interpreted
--- in the case where $n$ is timelike --- as being cotangent
to the worldline of observer families, i.e., the dual reference
frame relative velocity associated with such observers.

Denote $\{e_{a}\}$ ($a = 0,1,\ldots,D$) a basis for the tangent
space $T_x\Sigma$ at a point $x$ in $\Sigma$, and naturally the
cotangent space at $x\in\Sigma$ has an orthonormal basis $\{e^a\}$
such that $e^{a}(e_{b}) = \delta^{a}_{b}$. A reference frame at an
arbitrary point in the bulk is denoted by $\{e_{\alpha}\}$
  ($\alpha=0,1,2,\ldots,D+1$).  When a local coordinate chart is chosen, it is
possible to represent $e_{\alpha} = \partial/\partial x^{\alpha}
\equiv
\partial_{\alpha}$ and $e^{\alpha} = dx^{\alpha}$. The 1-form field orthogonal to the
sections of $T\Sigma$ --- the tangent bundle of $\Sigma$ --- can
now be written as $n = n_{\alpha} e^{\alpha}$,  and consider the
Gaussian coordinate $\ell$ orthogonal to the section of $T\Sigma$,
indicating how much an observer move out the $D$-dimensional brane
into the $(D+1)$-dimensional bulk.  A vector field $v = v^{\alpha}
e_{\alpha}$ in the bulk is split in components in the brane and
orthogonal to the brane, respectively as
 $v = v^{a} e_{a} + \ell e_{D+1}$. Since the bulk is endowed with a
non-degenerate bilinear symmetric form $g$ that can be written in
a coordinate basis as
 $g = g_{\alpha \beta}dx^{\alpha} \otimes dx^{\beta}$, the components of the metric in the brane and on the bulk are  denoted respectively by $q_{\alpha\beta}$ and
 $g_{\alpha\beta}$, and related by
\begin{equation}\label{neo}
g_{\alpha\beta} = q_{\alpha\beta} + n_{\alpha} n_{\beta}.
\end{equation}

The 1-form field $n$ orthogonal to $\Sigma$,  in the direction of increasing $\ell$ is
given by $n =  (\pa_{\alpha} \ell)\,e^{\alpha}$, and its covariant
components are explicitly given by $n_\alpha = \pa_{\alpha} \ell$.
Without loss of generality a timelike hypersurface $\Sigma$ is
taken, where a congruence of geodesics goes across it.  Denoting
the proper distance (or proper time) along
these geodesics by $\ell$, it is always possible to put $\ell=0$ on $\Sigma$. 

Denoting $\{x^\alpha\}$ a chart on both sides of the brane, define
another chart $\{y^a\}$ \emph{on} the brane. Here the same
notation used in \cite{PM} is adopted, where Latin indices is used for
hypersurface coordinates and Greek indices for coordinates in the
embedding spacetime. The brane can be parametrized by
$x^\alpha=x^\alpha(y^a)$,
and the terms $h^\alpha_a:=\frac{\pa x^\alpha}{\pa y^a} $ satisfy
$h^\alpha_a n_\alpha=0$. For displacements on the brane, it
follows that \beq g &=& g_{\alpha\beta}\,d x^\alpha \otimes d
x^\beta = g_{\alpha\beta}\,\Big(\frac{\pa x^\alpha}{\pa y^a}\,d
y^a \Big)\otimes
\Big(\frac{\pa x^\beta}{\pa y^b}\,d y^b \Big) \nonumber\\
&=& q_{ab}\,d y^a \otimes d y^b, \eeq and so the induced metric
components $q_{ab}$ on $\Sigma$ is related to $g_{\alpha\beta}$ by
$q_{ab} = g_{\alpha\beta}\,h^\alpha_a h^\beta_b.$

Denoting by $ [A] = \displaystyle\lim_{\ell\rightarrow 0^+}(A) -
\lim_{\ell\rightarrow 0^-}(A) $ the change in a differential form
field $A$ across the braneworld $\Sigma$ (wherein $\ell = 0$), the
continuity of the chart $x^\alpha$ and $\ell$ across $\Sigma$
implies that $n_\alpha$ and $h^\alpha_a$ are continuous, or,
equivalently,  $[n_\alpha]=[h^\alpha_a]=0$.

Now, using the Heaviside distribution $\Theta(\ell)$
properties\footnote{$\delta(\ell)$ is the Dirac distribution.}
\[
\Theta^2(\ell)=\Theta(\ell), \qquad \Theta(\ell)\Theta(-\ell)=0,
\qquad \frac{d}{d \ell}\,\Theta(\ell) = \delta(\ell),
\]
the metric components $g_{\alpha\beta}$ can be written as
distribution-valued tensor components
\[
g_{\alpha\beta}=\Theta(\ell)\,g^+_{\alpha\beta}+\Theta(-\ell)\,g^-_{\alpha\beta},
\]
where $g^+_{\alpha\beta}$ ($g^-_{\alpha\beta}$) denotes the metric
on the $\ell>0$ ($\ell<0$) side of $\Sigma$. Differentiating the
above expression, it reads
\[
\pa_\gamma g_{\alpha\beta}=\Theta(\ell)\,\pa_\gamma
g^+_{\alpha\beta}+\Theta(-\ell)\,\pa_\gamma g^-_{\alpha\beta} +
\delta(\ell)[g_{\alpha\beta}]n_\gamma.
\]
It can be shown that the condition $[g_{\alpha\beta}]=0$ must be
imposed for the connection to be defined as a
distribution\footnote{Basically, if the condition
 $[g_{\alpha\beta}]=0$ is not imposed, there appears the product
$\Theta\delta$, which is not well defined in the Levi-Civita part
of the connection. }, also implying  the `first' junction
condition $[h_{ab}]$ \cite{PM}.

Besides a curvature associated with the connection that endows the
bulk, in a Riemann-Cartan manifold the torsion associated with the
connection is in general non zero. Its components can be written
in terms of the connection components
$\Gamma^{\rho}{}_{\beta\alpha}$ as
\begin{equation}
T^{\rho}{}_{\alpha\beta} = \Gamma^{\rho}{}_{\beta\alpha} -
\Gamma^{\rho}{}_{\alpha\beta}. \label{tor}
\end{equation}
The general connection components is related to the Levi-Civita
connection components
${\stackrel{\circ}{\Gamma}}{}^{\rho}{}_{\alpha\beta}$ ---
associated with the spacetime metric $g_{\alpha\beta}$ components
--- through $\Gamma^{\rho}{}_{\alpha\beta} =
{\stackrel{\circ}{\Gamma}}{}^{\rho}{}_{\alpha\beta} +
K^{\rho}{}_{\alpha\beta}$, where $K^{\rho}{}_{\alpha \beta} =
\textstyle{\frac{1}{2}} \left( T_{\alpha}{}^{\rho}{}_{\beta} +
T_{\beta}{}^{\rho}{}_{\alpha} - T^{\rho}{}_{\alpha \beta} \right)$
denotes the contortion tensor components. It must be emphasized that
curvature and torsion are properties of a connection, not of
spacetime. For instance, the Christoffel and the general
connections present different curvature and torsion, although they
endow the very same manifold.

The easiest method of introducing torsion terms in the theory is via
the addition of an antisymmetric part in the affine connection.
The general connection components are related to the Levi-Civita
connection components $\mathring{\Gamma}{}^{\rho}{}_{\alpha\beta}$
--- associated with the spacetime metric $g_{\alpha\beta}$
components
--- through $\Gamma^{\rho}{}_{\alpha\beta} =
\mathring{\Gamma}{}^{\rho}{}_{\alpha\beta} +
K^{\rho}{}_{\alpha\beta}$, where $K^{\rho}{}_{\alpha \beta} =
\textstyle{\frac{1}{2}} \left( T_{\alpha}{}^{\rho}{}_{\beta} +
T_{\beta}{}^{\rho}{}_{\alpha} - T^{\rho}{}_{\alpha \beta} \right)$
denotes the contortion tensor components. Hereon the quantities
denoted by $\mathring{X}$ are constructed with the usual metric
compatible torsionless Levi-Civita connection components
$\mathring{\Gamma}{}^{\rho}{}_{\alpha\beta}$. We remark that the
source of contortion may be considered as the rank-2 antisymmetric
potential Kalb-Ramond (KR) field $B_{\alpha\beta}$, arising as a
massless mode in heterotic string theories \cite{cordas, SEN}.
Hereon we shall consider the formal geometric contortion, although
the contortion induced by the KR field can be considered in the
5-dimensional formalism when the prescription
$K^\rho_{\;\,\alpha\beta} = -
\frac{1}{M^{3/2}}H^\rho_{\;\,\alpha\beta}$ is taken into account,
where $H_{\rho\alpha\beta} = \partial_{[\rho}B_{\alpha\beta]}$ and
$M$ denotes the 5-dimensional Planck mass. The identification
between the KR field and the contortion can be always taken into
account when necessary, depending on the physical aspect of the
formalism that must be emphasized, although the formalism is not
precisely concerned with the fount of contortion, but with its
consequences.

Now the distribution-valued Riemann tensor is calculated, in order
to find the `second' junction condition  --- the Israel matching
condition. From the Christoffel symbols, it reads $
\Gamma^\alpha_{\;\,\beta\gamma} = \Theta(\ell)
\Gamma^{+\alpha}_{\;\;\,\beta\gamma}+\Theta(-\ell)\Gamma^{-\alpha}_{\;\;\,\beta\gamma},
$ where $\Gamma^{\pm\alpha}_{\;\;\,\beta\gamma}$ are the
Christoffel symbols obtained from $g^\pm_{\alpha\beta}$.  Thus
\[
\pa_\delta\Gamma^\alpha_{\;\,\beta\gamma} =
\Theta(\ell)\pa_\delta\Gamma^{+\alpha}_{\;\;\;\,\beta\gamma}
+\Theta(-\ell)\pa_\delta\Gamma^{-\alpha}_{\;\;\;\,\beta\gamma} +
\delta(\ell)[\Gamma^\alpha_{\;\,\beta\gamma}]n_\delta,
\]
and the Riemann tensor is given by $
R^\alpha_{\beta\gamma\delta}=\Theta(\ell)R^{+\alpha}_{\;\;\;\,\beta\gamma\delta}+\Theta(-\ell)R^{-\alpha}_{\;\;\;\,\beta\gamma\delta}
+\delta(\ell)A^\alpha_{\;\,\beta\gamma\delta}, $ where $
A^\alpha_{\;\,\beta\gamma\delta}=[\Gamma^\alpha_{\;\,\beta\delta}]n_\gamma-[\Gamma^\alpha_{\;\,\beta\gamma}]n_\delta$
\cite{PM}.

The next step is to find an explicit expression for the tensor
$A^\alpha_{\;\,\beta\gamma\delta}$. Observe that the continuity of the metric across $\Sigma$ implies
that the tangential derivatives of the metric must be also
continuous. If $\partial_\gamma g_{\alpha\beta} \equiv
g_{\alpha\beta, \gamma}$ is indeed discontinuous, this
discontinuity must be directed along the normal vector $n^\alpha$.
It is therefore possible to write
\[
[g_{\alpha\beta,\gamma}]=\kappa_{\alpha\beta}n_\gamma,
\]
for some tensor $\kappa_{\alpha\beta}$ (given explicitly by
$\kappa_{\alpha\beta}=[g_{\alpha\beta,\gamma}]n^\gamma$). Then
it follows that
\[
[{\stackrel{\circ}{\Gamma}}{}^\alpha_{\;\,\beta\gamma}] =
\me\,(\kappa^\alpha_{\;\,\beta}
n_\gamma+\kappa^\alpha_{\;\,\gamma} n_\beta - \kappa_{\beta\gamma}
n^\alpha),
\]
and supposing that the discontinuity in the torsion terms obey the
same rule as the discontinuity of $[g_{\alpha\beta,\gamma}]$, i.
e., that
$[T^{\alpha}_{\;\,\beta\gamma}]=\zeta^{\;\alpha}_{\beta}n_{\gamma}$,
it reads \beq [K^\alpha_{\;\,\beta\gamma}] &=&
\me\,(\zeta^{\;\,\alpha}_\beta n_\gamma+\zeta^{\;\,\alpha}_\gamma
n_\beta - \zeta^{\alpha}_{\;\,\beta} n_\gamma).\label{ca} \eeq The
components $\kappa_{\rho\sigma}$ emulate an intrinsic property of
the brane itself. The torsion is continuous along the brane, and
if there is some discontinuity, it is proportional to the extra
dimension. Such proportionality is given, in principle, by another
quantity  $\zeta^{\;\alpha}_{\beta}$ related to the brane. After
these considerations, it follows that \beq
[\Gamma^\alpha_{\;\,\beta\gamma}] &=&
 \me\,((\kappa^\alpha_{\;\,\beta} + \zeta^{\;\,\alpha}_\beta - \zeta^{\alpha}_{\;\,\beta}) n_\gamma
 +(\kappa^\alpha_{\;\,\gamma}+\zeta^{\;\,\alpha}_\gamma) n_\beta - \kappa_{\beta\gamma}
 n^\alpha),\nonumber
\eeq and hence \ba
A^\alpha_{\;\,\beta\gamma\delta}&=&\left.\frac{1}{2}\,(\kappa^\alpha_{\;\,\delta}
n_\beta n_\gamma - \kappa^\alpha_{\;\,\gamma} n_\beta n_\delta -
\kappa_{\beta\delta} n^\alpha n_\gamma
+\kappa_{\beta\gamma}n^\alpha n_\delta)\right. \nonumber\\&&+
\left. \me(\zeta_\delta^{\;\,\alpha}n_\beta n_\gamma -
\zeta_\gamma^{\;\,\alpha}n_\beta n_\delta).\right. \ea Denoting
$\kappa = \kappa^\alpha_{\;\,\alpha}$ and $\zeta =
\zeta^\beta_{\;\,\beta}$, and suitably contracting two indices, it
reads
\ba A_{\beta\delta}&=&\left.\frac{1}{2}(\kappa^\alpha_{\;\,\delta}
n_\beta n_\alpha -\kappa n_\beta n_\delta - \kappa_{\beta\delta} +
\kappa_{\beta\alpha} n^\alpha n_\delta)\right.\nonumber
\\&+&\left. \me(\zeta_\delta^{\;\,\alpha}n_\beta n_\alpha - \zeta
n_\beta n_\delta),\right. \eeq and also \beq A =
g^{\beta\delta}A_{\beta\delta} = (\kappa_{\alpha\delta} n^\alpha
n^\delta - \kappa) + \me(\zeta_{\delta\alpha}n^\delta n^\alpha -
\zeta). \nonumber\eeq

The $\delta$-function part of the Einstein tensor
$G_{\alpha\beta}:=R_{\alpha\beta}-\me g_{\alpha\beta}R$ is given
by \beq \label{Seq}
S_{\beta\delta} &=& A_{\beta\delta}-\me g_{\beta\delta}A\nonumber\\
&=& \frac{1}{2}\,(\kappa_{\;\,\delta}^{\alpha}n_\beta n_\alpha -
\kappa n_\beta n_\delta  -
\kappa_{\beta\delta} + \kappa_{\beta\alpha}n^\alpha n_\delta \nonumber\\
&& - g_{\beta\delta}(\kappa_{\rho\sigma}n^\rho n^\sigma -
\kappa))+ \frac{1}{2}\,(\zeta_{\delta}^{\;\,\alpha}n_\beta
n_\alpha - \zeta
n_\beta n_\delta) \nonumber\\
&& -\frac{1}{4} g_{\beta\delta}(\zeta_{\rho\sigma}n^\rho n^\sigma
- \zeta)).\label{opa} \eeq On the other hand, the total
stress-energy tensor is of the form
\[
\pi^\mathrm{\,total}_{\alpha\beta}=\Theta(\ell)\pi^+_{\alpha\beta}+\Theta(-\ell)\pi^-_{\alpha\beta}+\delta(\ell)\pi_{\alpha\beta},
\]
where $\pi^+_{\alpha\beta}$ and $\pi^-_{\alpha\beta}$ represent
the bulk stress-energy in the regions where $\ell> 0$ and $\ell <
0$ respectively, while $\pi_{\alpha\beta}$ denotes the
stress-energy localized on the hypersurface $\Sigma$ itself. From
the Einstein equations, it follows that $\pi_{\alpha\beta}=(G_{N})^{-1}
S_{\alpha\beta}$.

Note that, since $\pi_{\alpha\beta}$ is tangent to the brane, it
follows that $\pi_{\alpha\beta}n^{\beta}=0$. However, from
Eq.(\ref{Seq}) the following equation \beq 4 G_{N}
\pi_{\alpha\beta}n^\beta&=&\frac{1}{2}(\zeta_{\rho\sigma}n^\rho
n^\sigma - \zeta)n_\alpha \nonumber\\
&=& -\frac{1}{2}\mk \zeta_{\rho\sigma}{q}^{\rho\sigma} n_\alpha,
\eeq is derived, which means that, in order to keep the
consistence of the formalism, one has to impose
$\zeta_{\rho\sigma}q^{\rho\sigma}=0$, and the last term of
Eq.(\ref{opa}) vanishes. Note that  $\pi_{\alpha\beta}$ can be
expressed by $\pi_{ab}=\pi_{\alpha\beta}h^\alpha_a h^{\beta}_b$,
just using the $h_a^\alpha$ vierbein introduced in the previous
Section. So, taking into account that
$\pi_{\alpha\beta}=(G_{N})^{-1} S_{\alpha\beta}$ and
Eq.(\ref{opa}), it reads \cite{PM} \beq 4 G_{N}
\pi_{ab}=-\kappa_{\alpha\beta} h^\alpha_a
h^\beta_b+q^{rs}\kappa_{\mu\nu}h^\mu_r h^\nu_s q_{ab}.\eeq
Finally, relating the jump in the extrinsic curvature to
$\kappa_{\rho\sigma}$, via the covariant derivative associated to
$q_{\alpha\beta}$, the following expression can be obtained  from
Eq.(\ref{ca}):  \ba
[\nabla_{\alpha}n_\beta]&=&\left.\frac{1}{2}(\kappa_{\alpha\beta}-\kappa_{\gamma\alpha}n_{\beta}n^{\gamma}-\kappa_{\gamma\beta}n_\alpha
n^\gamma)\right.\nonumber\\&+&\left.\frac{1}{2}(\zeta^{\;\,\gamma}_\alpha
n_\beta+\zeta^{\;\,\gamma}_\beta n_\alpha -
\zeta^{\gamma}_{\;\,\alpha} n_\beta)n_\gamma .\right.\ea However,
it is clear that this jump of the extrinsic curvature across the
brane, $[\nabla_{\alpha}n_\beta]\equiv [\Xi_{\alpha\beta}]$, can
be also decomposed in terms of $h_a^\alpha$ vectors, leading to
\beq [\Xi_{ab}]=\frac{1}{2}\kappa_{\alpha\beta}h^\alpha_a
h^\beta_b .\eeq Hence, after all, it follows that \beq 2G_{N}
\pi_{ab}=-[\Xi_{ab}]+[\Xi]q_{ab}.\label{vai}\eeq It means that the
second matching condition is absolutely the same that is valid
without any torsion term.

Once investigated the matching conditions in the presence of
torsion terms, and under the assumptions of discontinuity
across the brane, both the
junctions conditions are shown to be the same as the usual case
($\Gamma^{\rho}_{\;\,\alpha\beta}={\stackrel{\circ}{\Gamma}}{}^{\rho}_{\;\,\alpha\beta}$).
We remark that, since the covariant derivative changes by torsion,
the extrinsic curvature is also modified, and then the
conventional arguments point in the direction of some modification
in the matching conditions. However, it seems that the r\^ole of
torsion terms in the braneworld picture is restricted to the
geometric part of effective Einstein equation on the brane. More
explicitly, looking at the equation that relates the Einstein
equation in four dimensions with bulk quantities (see, for example
\cite{JP}) we have
\begin{eqnarray}
^{(4)}\!G_{\rho\sigma}&=&\left.\frac{2k_{5}^{2}}{3}\Bigg(T_{\alpha\beta}q_{\rho}^{\;\alpha}q_{\sigma}^{\;\beta}+(T_{\alpha\beta}
n^{\alpha}n^{\beta}-\frac{1}{4}T)q_{\rho\sigma}\Bigg)\right.\nonumber
\\&&+\left. \Xi\Xi_{\rho\sigma}-\Xi_{\rho}^{\;\alpha}\Xi_{\alpha\sigma}-
\frac{1}{2}q_{\rho\sigma}(\Xi^{2}-\Xi^{\alpha\beta}\Xi_{\alpha\beta})\right.
\nonumber
\\&&-\left.\;^{(4)}\!C^{\alpha}_{\;\beta\gamma\epsilon}n_{\alpha}n^{\gamma}q_{\rho}^{\;\beta}q_{\sigma}^{\;\epsilon},\right.\label{gde}
\end{eqnarray}
where $T_{\rho\sigma}$ denotes the energy-momentum tensor,
$\Xi_{\rho\sigma}=q_{\rho}^{\;\alpha}q_{\sigma}^{\;\beta}\nabla_{\alpha}n_{\beta}$
is the extrinsic curvature, $k_5$ denotes the 5-dimensional gravitational constant, and
$^{(5)}\!C^{\alpha}_{\;\beta\rho\sigma}$ denotes the Weyl tensor.
By restricting to quantities evaluated on the brane, or tending to
the brane, we see that the only way to get some contribution from
torsion terms is via the term $^{(4)}\!G_{\rho\sigma}$, and also
via the Weyl tensor. It does not
intervene in the extrinsic curvature tending to the brane.

\section{Torsion influence on the projected equations on the brane}

In order to explicit the influence of contorsion terms in the
projected equations on the brane, we shall to complete the GC
program, from five to four dimensions, to the case with torsion.
Note the by imposing the $\mathbb{Z}_{2}$-symmetry, the extrinsic
curvature reads
\ba\Xi^{+}_{\alpha\beta}=-\Xi^{-}_{\alpha\beta}=-2G_{N}\Big(\pi_{\alpha\beta}-\frac{q_{\alpha\beta}\pi^{\gamma}_{\gamma}}{4}\Big),\label{c1}\ea
in such way that Eq.(\ref{vai}) reads\footnote{Hereon, we remove
the $+$ and $-$ labels.}
\ba\Xi_{\alpha\beta}=-G_{N}\Big(\pi_{\alpha\beta}-\frac{q_{\alpha\beta}\pi^{\gamma}_{\gamma}}{4}\Big).\label{c2}\ea

Decomposing the stress-tensor associated with the
bulk in
$T_{\alpha\beta}=-\Lambda g_{\alpha\beta}+\delta S_{\alpha\beta}$
and $S_{\alpha\beta}=-\lambda q_{\alpha\beta}+\pi_{\alpha\beta}$,
where $\Lambda$ is the bulk cosmological constant and $\lambda$
the brane tension, and substituting into Eq.(\ref{gde}) it follows
after some algebra\footnote{See, please, reference \cite{JP} for
all the details.}, \ba
^{(4)}\!G_{\mu\nu}=-\Lambda_{4}q_{\mu\nu}+8\pi
G_{N}\pi_{\mu\nu}+k_{5}^{4}Y_{\mu\nu}-E_{\mu\nu} ,\label{c3}\ea
where
$E_{\mu\nu}=^{(5)}\!C^{\alpha}_{\beta\gamma\sigma}n_{\alpha}n^{\gamma}q_{\mu}^{\beta}q_{\nu}^{\sigma}$
encodes the Weyl tensor contribution, $G_{N}=\frac{\lambda
k_{5}^{4}}{48\pi}$ is the analogous of the Newton gravitational
constant, the tensor $Y_{\mu\nu}$ is quadratic in the brane
stress-tensor and given by
$Y_{\mu\nu}=-\frac{1}{4}\pi_{\mu\alpha}\pi_{\nu}^{\alpha}+\frac{1}{12}\pi^{\gamma}_{\gamma}\pi_{\mu\nu}+\frac{1}{8}
q_{\mu\nu}\pi_{\alpha\beta}\pi^{\alpha\beta}-\frac{1}{2}q_{\mu\nu}(\pi^{\gamma}_{\gamma})^{2}$
and
$\Lambda_{4}=\frac{k_{5}^{2}}{2}\Big(\Lambda+\frac{1}{6}k_{5}^{2}\lambda^{2}\Big)$
is the effective brane cosmological constant.

It is well known that the Riemann and Ricci
tensors, and the curvature scalar written in terms of torsion are
related with their partners, constructed with the usual metric
compatible Levi-Civita connection by
\begin{eqnarray}
R^{\lambda}_{\;\;\tau\alpha\beta}&=&\left.\mathring{R}^{\lambda}_{\;\;\tau\alpha\beta}+\nabla_{\alpha}K^{\lambda}_{\;\;\tau\beta}
-\nabla_{\beta}K^{\lambda}_{\;\;\tau\alpha}+ K^{\lambda}_{\;\;\gamma\alpha}K^{\gamma}_{\;\;\tau\beta}
-K^{\lambda}_{\;\;\gamma\beta}K^{\gamma}_{\;\;\tau\alpha},\right.
\label{c4}
\end{eqnarray}
\begin{eqnarray}
R_{\tau\beta}&=&\mathring{R}_{\tau\beta}+\nabla_{\lambda}K^{\lambda}_{\;\;\tau\beta}-\nabla_{\beta}K^{\lambda}_{\;\;\tau\lambda}
+K^{\lambda}_{\;\;\gamma\lambda}K^{\gamma}_{\;\;\tau\beta}
-K^{\lambda}_{\;\;\tau\gamma}K^{\gamma}_{\;\;\lambda\beta}
\label{c5}
\end{eqnarray}
and
\begin{eqnarray}
R=\mathring{R}+2\nabla^{\lambda}K^{\tau}_{\;\;\lambda\tau}-K_{\tau\lambda}^{\;\;\;\lambda}K^{\tau
\lambda}_{\;\;\;\lambda}+K_{\tau\gamma\lambda}K^{\tau\lambda\gamma},\label{c6}
\end{eqnarray}
where the quantities $\mathring{X}$ are constructed with the usual
metric compatible Levi-Civita connection, and $\nabla$ denotes the
covariant derivative {\it without} torsion. Clearly such relations
holds in any dimension. Therefore, by denoting $D_{\mu}$ the
covariant 4-dimensional derivative acting on the brane, it is easy
to see that, from Eqs.(\ref{c4}),(\ref{c5}), and (\ref{c6}), the
Einstein tensor on the brane is given by
\begin{eqnarray}
^{(4)}\!G_{\mu\nu}&=&\left.
^{(4)}\!\!\mathring{G}_{\mu\nu}+D_{\lambda}\;^{(4)}\!K^{\lambda}_{\;\;\mu\nu}-D_{\nu}\;^{(4)}\!K^{\lambda}_{\;\;\mu\lambda}+ ^{(4)}\!K^{\lambda}_{\;\;\gamma\lambda}\;^{(4)}K^{\gamma}_{\;\;\mu\nu}
-\;^{(4)}K^{\lambda}_{\;\;\mu\gamma}\;^{(4)}K^{\gamma}_{\;\;\lambda\nu}-q_{\mu\nu}\Big(
D^{\lambda}\;^{(4)}\!K^{\tau}_{\;\;\lambda\tau}+\frac{1}{2}\;^{(4)}\!K^{\;\;\;\lambda}_{\tau\lambda}
\;^{(4)}\!K^{\tau\gamma}_{\;\;\;\gamma}\right.\nonumber\\&&\left.+\frac{1}{2}\;^{(4)}\!K_{\tau\gamma\lambda}\;^{(4)}K^{\tau\gamma\lambda}\Big)\right..
\label{c7}
\end{eqnarray}
Note the appearance of terms multiplying the brane metric. As it
shall be seen, these terms compose a new effective cosmological
constant.

The presence of extra dimensions seems to be an almost inescapable
characteristic of high-energy physics based upon the auspices of
string theory. In this context, specific string theory inspired
scenarios, in which our universe is modeled by a brane --- the
braneworld scenario --- acquired special attention \cite{RS} due
to the possibility of solving the hierarchy problem.
Concomitantly, the presence of torsion is also an output of string
theory \cite{cordas}. Indeed, when gravitation is recovered from
string theory, a plenty of physical fields abound, including the
torsion field. In this context, among other motivations, it seems
natural to explore some properties of braneworld models in the
presence of torsion.

\section{Measurable Torsion Effects}

In the previous Section we proved that although the
presence of torsion terms in the connection does not modify the
Israel-Darmois matching conditions. The factors involving contortion alter drastically the effective
Einstein equation on the brane, and also the function involving contortion
terms that is analogous to the effective
cosmological constant as well.

We shall use such results to extend the bulk metric Taylor expansion in terms of the brane metric, in a direction
orthogonal to the brane, encompassing torsion terms. As an
immediate application, the corrections in a black hole horizon
area due to contortion terms are achieved.

Using the Einstein tensor on the brane encoding torsion terms, the
$E_{\mu\nu}$ tensor can be expressed in terms of the bulk
contortion terms by
\begin{eqnarray}
\hspace{-.5cm}E_{\kappa\delta}&=&\left.\mathring{E}_{\kappa\delta}+\Big(\nabla_{[\nu}K^{\mu}_{\;\;\alpha\beta]}+
K^{\mu}_{\;\;\gamma[\nu}K^{\gamma}_{\;\;\alpha\beta]}\Big)n_{\mu}n^{\nu}
q_{\kappa}^{\alpha}q_{\delta}^{\beta}\right.-\frac{2}{3}(q_{\kappa}^{\alpha}q_{\delta}^{\beta}+n^{\alpha}n^{\beta}q_{\kappa\delta})
\Big(\nabla_{[\lambda}K^{\lambda}_{\;\;\beta\alpha]}+K^{\lambda}_{\;\;\gamma\lambda}K^{\gamma}_{\;\;\beta\alpha}
-K^{\sigma}_{\;\;\beta\gamma}K^{\gamma}_{\;\;\sigma\alpha}\Big)\nonumber\\&&+\frac{1}{6}q_{\kappa\delta}\left.
\Big(2\nabla^{\lambda}K^{\tau}_{\;\;\lambda\tau}-K_{\tau\lambda}^{\;\;\;\lambda}K^{\tau\gamma}_{\;\;\;\gamma}+K_{\tau\gamma\lambda}
K^{\tau\lambda\gamma} \Big), \right. \label{c8}
\end{eqnarray} where $\nabla_{\mu}$ is the bulk covariant
derivative. Now, the explicit influence of the contortion terms in
the Einstein brane equation can be visualized. From
Eqs.(\ref{c3}), (\ref{c8}) and expressing the torsion terms of the
Einstein brane tensor (see Eq. (20) of reference \cite{NOIS}), it
follows that
\begin{widetext}
\begin{eqnarray}&&
\left.
\hspace{-1cm}^{(4)}\!\mathring{G}_{\mu\nu}+D_{[\lambda}\;^{(4)}\!K^{\lambda}_{\;\;\mu\nu]}+
^{(4)}\!K^{\delta}_{\;\;\gamma\delta}\;^{(4)}\!K^{\lambda}_{\;\;\mu\nu}
-
\;^{(4)}\!K^{\sigma}_{\;\;\nu\gamma}\;^{(4)}\!K^{\gamma}_{\;\;\sigma\mu}
=-\tilde{\Lambda}_{4}q_{\mu\nu}+8\pi
G_{N}\pi_{\mu\nu}+k_{5}^{4}Y_{\mu\nu}-\mathring{E}_{\mu\nu}\right.\nonumber
\\&&\hspace{-1cm}+q_{\mu}^{\alpha}q_{\nu}^{\beta} \left.\Bigg[\frac{2}{3}\Big(\nabla_{[\lambda}
K^{\lambda}_{\;\;\beta\alpha]}+K^{\sigma}_{\;\;\gamma\sigma}K^{\gamma}_{\;\;\beta\alpha}
-K^{\lambda}_{\;\;\beta\gamma}K^{\gamma}_{\;\;\lambda\alpha}\Big)-n_{\rho}n^{\sigma}\Big(\nabla_{[\sigma}K^{\rho}_{\;\;\alpha\beta]}+K^{\rho}_{\;\;\gamma[\sigma}K^{\gamma}_{\;\;\alpha\beta]}
\Big)\Bigg], \right. \label{c9}
\end{eqnarray}
where
\begin{eqnarray}
\tilde{\Lambda}_{4}&\equiv&\left.
\Lambda_{4}-D^{\lambda}\;^{(4)}\!K^{\tau}_{\;\;\lambda\tau}+\frac{1}{2}\;^{(4)}K_{\tau\alpha}^{\;\;\;\alpha}\;^{(4)}K^{\tau\lambda}_{\;\;\;\lambda}
-\frac{1}{2}\;\;^{(4)}K_{\tau\gamma\lambda}\;^{(4)}K^{\tau\lambda\gamma}-\frac{2}{3}n^{\alpha}n^{\beta}\Big(\nabla_{\lambda}K^{\lambda}_{\;\;\beta\alpha}-
\nabla_{\alpha}K^{\lambda}_{\;\;\beta\lambda}\right.\nonumber\\&&+\left.K^{\lambda}_{\;\;\gamma\lambda}K^{\gamma}_{\;\;\beta\alpha}
-K^{\sigma}_{\;\;\beta\gamma}K^{\gamma}_{\;\;\sigma\alpha}\Big)+\frac{1}{6}\Big(2\nabla^{\lambda}K^{\tau}_{\;\;\lambda\tau}-
K_{\tau\alpha}^{\;\;\;\alpha}K^{\tau\lambda}_{\;\;\;\lambda}+K_{\tau\gamma\lambda}K^{\tau\lambda\gamma}\Big).\label{c10}
\right.
\end{eqnarray}
\end{widetext} The function $\tilde{\Lambda}_4$ above is usually called
 effective cosmological ``constant'' in the literature, in the sense that all its terms are multiplied by the brane metric
 in the Einstein effective equation (\ref{c9}). Eqs. (\ref{c9}) and (\ref{c10}) show that the factors involving
contortion, both in four and in five dimensions, modify
drastically the effective Einstein equation on the brane and the
effective cosmological constant as well.

Now, let us look at some deviations of the black hole horizon
coming from the bulk torsion terms. Hereon in this Section we
assume vacuum on the brane ($\pi_{\mu\nu}=0=Y_{\mu\nu}$) and
neglect the contribution of the effective cosmological constant
term, which is expected to be smaller, by some orders of
magnitude, than the contribution of the term Weyl \cite{Maartens}.
Using a Taylor expansion in the extra dimension in order to probe
properties of a static black hole on the brane \cite{Dad}, the
bulk metric can be written as
\begin{eqnarray}\label{metrica}
g_{\mu\nu}(x,y) &=& q_{\mu\nu} - (\mathring{E}_{\mu\nu} +
A_{\mu\nu} )y^2 - \frac{2}{l}(\mathring{E}_{\mu\nu} + A_{\mu\nu})
y^3 +\frac{1}{12}\left.\Bigg(\left({\Box}\mathring{E}_{\mu\nu} -
\frac{32}{l^2}\mathring{E}_{\mu\nu} +
2\mathring{R}_{\mu\alpha\nu\beta}\mathring{E}^{\alpha\beta} +
6\mathring{E}_{\mu}^{\;
\alpha}\mathring{E}_{\alpha\nu}\right)\right.\nonumber\\&&+
\left.\left({\Box}{A}_{\mu\nu} - \frac{32}{l^2}{A}_{\mu\nu} + 2(
\nabla_{[\nu}K_{\mu\alpha\beta]}){A}^{\alpha\beta} +
2K_{\mu\gamma[\nu}K^{\gamma}_{\;\;\alpha\beta]}{A}^{\alpha\beta} +
6\mathring{A}_{\mu}^{\; \alpha}\mathring{A}_{\alpha\nu}\right)
 \Bigg)\right. y^4 + \cdots
\nonumber\end{eqnarray} \noindent where \begin{eqnarray}A_{\mu\nu}
&=&  \left.\Big(\nabla_{[\delta}K^{\kappa}_{\;\;\alpha\beta]}+
K^{\kappa}_{\;\;\gamma[\beta}K^{\gamma}_{\;\;\alpha\delta]}\Big)n_{\kappa}n^{\delta}
q_{\mu}^{\alpha}q_{\nu}^{\beta}\right. +\frac{1}{6}q_{\mu\nu}
\Big(2\nabla^{\lambda}K^{\tau}_{\;\;\lambda\tau}-K_{\tau\lambda}^{\;\;\;\lambda}K^{\tau\gamma}_{\;\;\;\gamma}+K_{\tau\gamma\lambda}
K^{\tau\lambda\gamma}
\Big)\nonumber\\&&\left.-\frac{2}{3}(q_{\mu}^{\alpha}q_{\nu}^{\beta}+n^{\alpha}n^{\beta}q_{\mu\nu})
\Big(\nabla_{[\lambda}K^{\lambda}_{\;\;\beta\alpha]}+K^{\lambda}_{\;\;\gamma\lambda}K^{\gamma}_{\;\;\beta\alpha}
-K^{\sigma}_{\;\;\beta\gamma}K^{\gamma}_{\;\;\sigma\alpha}\Big)\right.\nonumber\label{amunu}\end{eqnarray}
and $\Box$ denotes the usual d'Alembertian. As in \cite{Maartens},
it shows in particular that the propagating effect of $5D$ gravity
arises only at the fourth order of the expansion. For a static
spherical metric on the brane given by \begin{equation}\label{124}
g_{\mu\nu}dx^{\mu}dx^{\nu} = - F(r)dt^2 + \frac{dr^2}{H(r)} +
r^2d\Omega^2,
\end{equation}
\noindent
 the projected Weyl term on the brane is given by the expressions\footnote{In the three expressions below, the indices $r$ and $\theta$ strictly denote
 the coordinates, and can not be confounded with summation indices.}
 \begin{eqnarray}
E_{00} &=& \frac{F}{r}\left(H' - \frac{1 - H}{r}\right) +
\Big(\nabla_{\nu}K^{\mu 00}-\nabla_{0}K^{\mu}_{\;\;0\nu}+
K^{\mu}_{\;\;\gamma\nu}K^{\gamma}_{\;\;00}-K^{\mu}_{\;\;\gamma
0}K^{\gamma}_{\;\;0\nu}\Big)n_{\mu}n^{\nu}
 F^2\nonumber\\&&-\frac{2}{3} F(F-1)
\Big(\nabla_{\lambda}K^{\lambda}_{\;\;00}-\nabla_{0}K^{\lambda}_{\;\;0\lambda}+K^{\lambda}_{\;\;\gamma\lambda}K^{\gamma}_{\;\;00}
-K^{\sigma}_{\;\;0\gamma}K^{\gamma}_{\;\;\sigma 0}\Big)
+\frac{1}{6} F
\Big(2\nabla^{\lambda}K^{\tau}_{\;\;\lambda\tau}-K_{\tau\lambda}^{\;\;\;\lambda}K^{\tau\gamma}_{\;\;\;\gamma}+K_{\tau\gamma\lambda}
K^{\tau\lambda\gamma} \Big),\nonumber\\
E_{rr} &=& -\frac{1}{rH}\left(\frac{F'}{F} - \frac{1 -
H}{r}\right)  + \Big(\nabla_{\nu}K^{\mu
rr}-\nabla_{r}K^{\mu}_{\;\;r\nu}+
K^{\mu}_{\;\;\gamma\nu}K^{\gamma}_{\;\;rr}-K^{\mu}_{\;\;\gamma
r}K^{\gamma}_{\;\;r\nu}\Big)n^{\mu}n^{\nu}
 H^{-2}\nonumber\\&&-\frac{2}{3} H^{-1}(H^{-1}-(n^r)^2)
\Big(\nabla_{\lambda}K^{\lambda}_{\;\;rr}-\nabla_{r}K^{\lambda}_{\;\;r\lambda}+K^{\lambda}_{\;\;\gamma\lambda}K^{\gamma}_{\;\;rr}
-K^{\sigma}_{\;\;r\gamma}K^{\gamma}_{\;\;\sigma
r}\Big)\nonumber\\&& +\frac{1}{6} H^{-1}\left.
\Big(2\nabla^{\lambda}K^{\tau}_{\;\;\lambda\tau}-K_{\tau\lambda}^{\;\;\;\lambda}K^{\tau\gamma}_{\;\;\;\gamma}+K_{\tau\gamma\lambda}
K^{\tau\lambda\gamma} \Big) \right. ,\nonumber\\
E_{\theta\theta} &=& -1 + H +\frac{r}{2}H\left(\frac{F'}{F} +
\frac{H'}{H}\right) + \Big(\nabla_{\nu}K^{\mu
\theta\theta}-\nabla_{\theta}K^{\mu}_{\;\;\theta\nu}+
K^{\mu}_{\;\;\gamma\nu}K^{\gamma}_{\;\;\theta\theta}-K^{\mu}_{\;\;\gamma
\theta}K^{\gamma}_{\;\;\theta\nu}\Big)n_{\mu}n^{\nu}
 r^4\nonumber\\&&-\frac{2}{3} r^2(r^2+1)
\Big(\nabla_{\lambda}K^{\lambda}_{\;\;\theta\theta}-\nabla_{\theta}K^{\lambda}_{\;\;\theta\lambda}+K^{\lambda}_{\;\;\gamma\lambda}K^{\gamma}_{\;\;\theta\theta}
-K^{\sigma}_{\;\;\theta\gamma}K^{\gamma}_{\;\;\sigma \theta} -
\frac{1}{2} \nabla^{\lambda}K^{\tau}_{\;\;\lambda\tau}+\frac{1}{4}
K_{\tau\lambda}^{\;\;\;\lambda}K^{\tau\gamma}_{\;\;\;\gamma}-\frac{1}{4}K_{\tau\gamma\lambda}
K^{\tau\lambda\gamma} \Big).\label{1333}
\end{eqnarray}

\noindent Note that in Eq.(\ref{124}) the metric reduces to the
Schwarzschild one, if $F(r)$ equals $H(r)$. The exact
determination of these radial functions remains an open problem in
black hole theory on the brane \cite{Maartens}.

These components allow one to evaluate the metric coefficients in
Eq.(\ref{metrica}). The area of the $5D$ horizon is determined by
$g_{\theta\theta}$. Defining $\psi(r)$ as the deviation from a
Schwarzschild form $H$, i.e.,
\begin{equation}\label{h}
H(r) = 1 - \frac{2M}{r} + \psi(r),
\end{equation}
\noindent where $M$ is constant, yields
\begin{eqnarray}\label{gtheta}
g_{\theta\theta}(r,y) &=& r^2  + \psi'\left(1 +
\frac{2}{l}y\right) + \Big(\nabla_{\nu}K^{\mu
\theta\theta}-\nabla_{\theta}K^{\mu}_{\;\;\theta\nu}+
K^{\mu}_{\;\;\gamma\nu}K^{\gamma}_{\;\;\theta\theta}-K^{\mu}_{\;\;\gamma
\theta}K^{\gamma}_{\;\;\theta\nu}\Big)n_{\mu}n^{\nu}
 r^4\nonumber\\&-&\frac{2}{3} r^2(r^2+1)
\Big(\nabla_{[\lambda}K^{\lambda}_{\;\;\theta\theta]}+K^{\lambda}_{\;\;\gamma\lambda}K^{\gamma}_{\;\;\theta\theta}
-K^{\sigma}_{\;\;\theta\gamma}K^{\gamma}_{\;\;\sigma \theta}
-\frac{1}{2}
\nabla^{\lambda}K^{\tau}_{\;\;\lambda\tau}+\frac{1}{4}K_{\tau\lambda}^{\;\;\;\lambda}K^{\tau\gamma}_{\;\;\;\gamma}-\frac{1}{4}K_{\tau\gamma\lambda}
K^{\tau\lambda\gamma} \Big)y^2\nonumber\\ &+&\cdots \label{modif}
\end{eqnarray}
\noindent It shows how $\psi$ \emph{and} the contortion and its
derivatives determine the variation in the area of the horizon along
the extra dimension. Also, the variation in the black string
properties can be extracted. Obviously, when the torsion goes to
zero, all the results above are led to the ones obtained in
\cite{Maartens}, \cite{Dad}, and references therein.
 In particular, Eq.(\ref{metrica}) --- when the torsion, and consequently $A_{\mu\nu}$ defined in Eq.(\ref{amunu}),  goes to zero ---
 is led to the results previously obtained in \cite{Maartens}.

As the area of the a black hole $5D$ horizon is determined by
$g_{\theta\theta}$, in particular it may indicate observable
signatures of corrections induced by contortion terms, since for a
given fixed effective extra dimension size, supermassive black
holes give the upper limit of variation in luminosity of quasars.
Also, it is possible to re-analyze how the quasar luminosity
variation behaves as a function of the AdS$_5$ bulk radius ---
corrected by contortion terms --- in some solar mass range, as in
\cite{merc} and references therein.

Furthermore,  braneworld measurable corrections induced by
contortion terms for  quasars, associated with Schwarzschild and
Kerr black holes, by their luminosity observation are important.
These corrections in a torsionless context were shown to be more
notorious for mini-black holes, where the Reissner-Nordstrom
radius in a braneworld scenario is shown to be around a hundred
times bigger than the standard Reissner–Nordstrom radius
associated with mini-black holes, besides mini-black holes being much more
sensitive to braneworld effects. It is possible to repeat all the
comprehensive and computational procedure in \cite{merc} in order
to verify how the contortion effects in Eq.(\ref{modif}) can
modify even more the above-mentioned results.

The modification in the area of the black hole horizon due to
torsion terms, whose functional form is depicted in Eq.
(\ref{modif}), can be better appreciated in a specific basis, i.
e., an explicit {\it ansatz} for the spacetime metric. This is,
however, out of the scope of the present work. The important point
here is that torsion terms do affect the black hole horizon and
the departure from the usual (torsionless) case is precisely given
by Eq. (\ref{modif}). In the next Section we extend and apply the
braneworld sum rules to the case with torsion, considering some
estimates of the torsion effects.

\section{Sum rules with torsion}

In this Section we shall derive the consistency conditions for
braneworld scenarios embedded in a Riemann-Cartan manifold. The general
procedure is quite similar to the one found in \cite{RK,LE} and we
shall comprise some of the general formulation here, for the sake of
completeness.

We start in a very general setup, analyzing a $D$-dimensional bulk
spacetime geometry, endowed with a non-factorizable metric \ba
ds^{2}&=&\left.G_{AB}dX^{A}dX^{B}\right.\nonumber\\&=&
\left.W^{2}(r)g_{\alpha\beta}dx^{\alpha}dx^{\beta}+g_{ab}(r)dr^{a}dr^{b}\label{2},\right.
\ea where $W^{2}(r)$ is the warp factor, $X^{A}$ denotes the
coordinates of the full $D$-dimensional bulk, $x^{\alpha}$ stands
for the $(p+1)$ non-compact spacetime coordinates, and $r^{a}$
labels the $(D-p-1)$ directions in the internal compact space. The
$D$-dimensional Ricci tensor can be related to its lower
dimensional partners by \cite{RK} \begin{eqnarray}
R_{\mu\nu}&=&\bar{R}_{\mu\nu}-\frac{g_{\mu\nu}}{(p+1)W^{p-1}}\nabla^{2}W^{p+1},\label{21}
\\
R_{ab}&=&\tilde{R}_{ab}-\frac{p+1}{W}\nabla_{a}\nabla_{b}W,\label{3}\end{eqnarray}
where $\tilde{R}_{ab}$, $\nabla_{a}$ and $\nabla^{2}$ are
respectively the Ricci tensor, the covariant derivative, and the
Laplacian operator constructed by means of  the internal space
metric $g_{ab}$. $\bar{R}_{\mu\nu}$ is the Ricci tensor derived
from $g_{\mu\nu}$. Denoting the three curvature scalars by
$R=G^{AB}R_{AB}$, $\bar{R}=g^{\mu\nu}\bar{R}_{\mu\nu}$, and
$\tilde{R}=g^{ab}\tilde{R}_{ab}$ we have, from Eqs.(\ref{2}) and
(\ref{3}), \ba
\frac{1}{p+1}\Big(W^{-2}\bar{R}-R^{\mu}_{\;\,\mu}\Big)=pW^{-2}\nabla
W\cdot\nabla W+W^{-1}\nabla^{2}W \label{4}\ea and \be
\frac{1}{p+1}\Big(\tilde{R}-R_{\;\,a}^{a}\Big)=W^{-1}\nabla^{2}W,\label{5}
\ee where $R^{\mu}_{\;\,\mu}\equiv W^{-2}g^{\mu\nu}R_{\mu\nu}$ and
$R^{a}_{\;\,a}\equiv g^{ab}R_{ab}$
($R=R^{\mu}_{\;\,\mu}+R^{a}_{\;\,a}$). It can be easily verified
that for an arbitrary constant $\xi$ the following identity holds
\be \frac{\nabla \cdot (W^{\xi}\nabla W)}{W^{\xi+1}}=\xi
W^{-2}\nabla W\cdot \nabla W+W^{-1}\nabla^{2}W \label{6}.\ee
Combining the above equation with Eqs.(\ref{2}) and (\ref{3}) we
have \ba \nabla \cdot (W^{\xi}\nabla
W)=\frac{W^{\xi+1}}{p(p+1)}[\xi\big(W^{-2}\bar{R}-R^{\mu}_{\;\,\mu}\big)
+(p-\xi)\big(\tilde{R}-R^{a}_{\;\,a}\big)].\label{7} \ea

The $D$-dimensional Einstein equation is given by \be R_{AB}=8\pi
G_{D}\Big(T_{AB}-\frac{1}{D-2}G_{AB}T\Big),\label{8} \ee where
$G_{D}$ is the gravitational constant in $D$ dimensions. It is
easy to write down the following equations: \ba
R^{\mu}_{\;\,\mu}=\frac{8\pi
G_{D}}{D-2}(T^{\mu}_{\;\,\mu}(D-p-3)-T^{m}_{\;\,m}(p+1)),\quad\qquad
R^{m}_{\;\,m}=\frac{8\pi
G_{D}}{D-2}(T^{m}_{\;\,m}(p-1)-T^{\mu}_{\;\,\mu}(D-p-1)).\label{9}\ea
In the above equations we set
$T^{\mu}_{\;\,\mu}=W^{-2}g_{\mu\nu}T^{\mu\nu}$
($T^{M}_{\;\,M}=T^{\mu}_{\;\,\mu}+T^{m}_{\;\,m}$). Now, it is
possible to relate $R^{\mu}_{\;\,\mu}$ and $R^{m}_{\;\,m}$ in
Eq.(\ref{7}) in terms of the stress-tensor. Note that the left
hand side of  Eq.(\ref{7}) vanishes upon integration
along a compact internal space. Hence, taking all that into
account we have \ba  \oint
W^{\xi+1}\Bigg(T^{\mu}_{\;\,\mu}[(p-2\xi)(D-p-1)+2\xi]+T^{m}_{\;\,m}p\,(2\xi-p+1)+\frac{D-2}{8\pi
G_{D}}[(p-\xi)\tilde{R}+\xi \bar{R}W^{-2}]\Bigg)=0.\label{11} \ea

Let us to particularize the analysis for a 5-dimensional bulk,
since it describes the phenomenologically interesting case.
Besides, it makes the conclusions obtained here applicable to the
case studied in \cite{NOIS}, in continuity to the program of
developing formal concepts to braneworld scenarios with torsion. In this
way $D=5$, $p=3$, and consequently $\tilde{R}=0$, because there is
just one dimension on the internal space. With such specifications
and assuming implicitly, as usual, that the brane action volume element
 does not depend on torsion\footnote{In this way, we
guarantee that the brane volume element reduces to $d^{4}x$ in the
limit of null torsion and flat space.}, Eq.(\ref{11}) becomes \ba
\oint
W^{\xi+1}\Bigg(T^{\mu}_{\;\;\mu}+2(\xi-1)T^{m}_{\;\;m}+\frac{\xi}{\kappa_{5}^{2}}\bar{R}W^{-2}\Bigg)=0,\label{12}
\ea where $8\pi G_{5}=\kappa_{5}^{2}=\frac{8\pi}{M_{5}^{3}}$, with
$M_{5}$ denoting the 5-dimensional Planck mass. In order to
implement torsion terms in our analysis, the expressions for the
Riemann and Ricci tensors  in terms of contortion components
related with their partners --- constructed with the usual metric
compatible Levi-Civita connection

\begin{eqnarray}
\bar{R}^{\lambda}_{\;\;\tau\alpha\beta}&=&\mathring{\bar{R}}^{\lambda}_{\;\tau\alpha\beta}+\nabla_{[\alpha}{}^{\scriptstyle{(4)}}K^{\lambda}_{\;\;\tau\beta]}
+{}^{\scriptstyle{(4)}}K^{\lambda}_{\;\;\gamma[\alpha}{}^{\scriptstyle{(4)}}K^{\gamma}_{\;\;\tau\beta]}\nonumber\\
\bar{R}_{\tau\beta}&=&\mathring{\bar{R}}_{\tau\beta}+\nabla_{[\lambda}{}^{\scriptstyle{(4)}}K^{\lambda}_{\;\;\tau\beta]}+{}^{\scriptstyle{(4)}}K^{\lambda}_{\;\;\gamma\lambda}{}^{\scriptstyle{(4)}}K^{\gamma}_{\;\;\tau\beta}
-{}^{\scriptstyle{(4)}}K^{\lambda}_{\;\;\tau\gamma}{}^{\scriptstyle{(4)}}K^{\gamma}_{\;\;\lambda\beta}
\label{c51}
\end{eqnarray}
where the label ``${\scriptstyle{(4)}}$'' on the contortion terms denotes the
contortion of the 3-branes and the covariant derivative is
considered when  a connection that
presents {\it no} torsion is taken into account.
First, however, note that in order to reproduce the observable
universe one can put $\mathring{\bar{R}}=0$ with $10^{-120}M_{Pl}$
of confidence level, where $M_{Pl}$ is the Planck mass. Note
that the observations concerning the scalar curvature are related
to the torsionless $\mathring{\bar{R}}$, not to $\bar{R}$. So,
taking it into account it follows that \be \oint
W^{\xi+1}\Bigg[T^{\mu}_{\;\,\mu}+2(\xi-1)T^{m}_{\;\;m}+\frac{\xi
W^{-2}}{\kappa_{5}^{2}}\Big(2D^{\lambda}\,^{(4)}\!K^{\tau}_{\;\lambda\tau}-
^{(4)}\!\!K_{\tau\lambda}^{\;\;\;\lambda}\,^{(4)}\!K^{\tau\lambda}_{\;\;\;\lambda}+
^{(4)}\!\!K_{\tau\gamma\lambda}\,^{(4)}\!K^{\tau\lambda\gamma}
\Big)\Bigg]=0.\label{14}\ee

In order to proceed with the consistency conditions we specify the
standard {\it ansatz} for the stress-tensor. Assuming that there
are no other types of matter in the bulk, except the branes and
the cosmological constant, we have \cite{LE} \ba
T_{MN}=-\frac{\Lambda}{\kappa_{5}^{2}}G_{MN}-\sum_{i}T_{3}^{(i)}P[G_{MN}]_{3}^{(i)}\delta(y-y_{i}),\label{15}
\ea where $\Lambda$ is the bulk cosmological constant,
$T_{3}^{(i)}$ is the tension associated to the i$^{th}$-brane and
$P[G_{MN}]_{3}^{(i)}$ is the pull-back of the metric to the
3-brane. The partial traces of (\ref{15}) are given by \ba
T_{\;\,\mu}^{\mu}=\frac{-4\Lambda}{\kappa_{5}^{2}}-4\sum_{i}T_{3}^{(i)}\delta(y-y_{i}),
\qquad \text{and}\qquad
T_{\;\,m}^{m}=-\frac{\Lambda}{\kappa_{5}^{2}},\label{16}\ea in
such way that  Eq.(\ref{14}) becomes \ba \oint
W^{\xi+1}\Bigg[\frac{2\Lambda}{\kappa_{5}^{2}}(\xi+1)+4\sum_{i}T_{3}^{(i)}\delta(y-y_{i})-\frac{\xi
W^{-2}}{\kappa_{5}^{2}}\Big(2D^{\lambda}\,^{(4)}\!K^{\tau}_{\;\lambda\tau}-
^{(4)}\!\!K_{\tau\lambda}^{\;\;\;\lambda}\,^{(4)}\!K^{\tau\lambda}_{\;\;\;\lambda}+
^{(4)}\!\!K_{\tau\gamma\lambda}\,^{(4)}\!K^{\tau\lambda\gamma}
\Big)\Bigg]=0.\label{18}\ea As one can see, this formalism can be
applied for a several branes scenario. The number of branes,
nevertheless, is not so important to our analysis. To fix ideas
let us particularize the formalism to the two branes case. Denoting
$T_{3}^{(1)}=\lambda$, the visible brane,
$T_{3}^{2}=\tilde{\lambda}$, and assuming that neither the
cosmological constant nor the branes contortion terms do depend on
the extra dimension, Eq.(\ref{18}) gives \ba 4\lambda
W_{\lambda}^{\xi+1}+4\tilde{\lambda}W_{\tilde{\lambda}}^{\xi+1}+\frac{2\Lambda}{\kappa_{5}^{2}}(\xi+1)\oint
W^{\xi+1}
-\frac{\xi}{\kappa_{5}^{2}}\Big(2D^{\lambda}\,^{(4)}\!K^{\tau}_{\;\lambda\tau}-
^{(4)}\!\!K_{\tau\lambda}^{\;\;\;\lambda}\,^{(4)}\!K^{\tau\lambda}_{\;\;\;\lambda}+
^{(4)}\!\!K_{\tau\gamma\lambda}\,^{(4)}\!K^{\tau\lambda\gamma}
\Big)\oint W^{\xi-1}=0,\label{19}\ea where
$W_{\lambda}=W(y=y_{1})$ and $W_{\tilde{\lambda}}=W(y=y_{2})$. Now
some physical outputs of the general Eq.(\ref{19}) are analyzed,
in order to investigate the viability of braneworld scenarios with torsion.
The first case we shall look at relates a factorizable geometry.
Nevertheless, before going forward, we shall emphasize that if one
implements the torsion null case in  Eq.(\ref{19}), it is easy
to see that for $\xi=-1$ one recovers the well known fine tuning
of the Randall-Sundrum model, i.e., \ba
\lambda+\tilde{\lambda}=0\label{20},\ea as expected.
\subsection{Non-warped compactifications with torsion}

The non-warped case is implemented by imposing $W=1$, working
then in a factorizable spacetime geometry. The general approach on
consistency conditions, as exposed before, allows this
possibility. In this Subsection we are therefore concerned with
the viability of braneworld scenarios in the general scope analyzed in
reference \cite{GVALI} and in the presence of torsion. The case we
are going to describe here is not the most interesting. We shall,
however, study a little further this simplified case, since it
can provide some physical insight to the warped case.

From Eq.(\ref{19}), the non-warped case reads \ba
\frac{2\Lambda}{\kappa_{5}^{2}}(\xi+1)V+4\lambda+4\tilde{\lambda}-\frac{\xi
V}{\kappa_{5}^{2}}\Big(2D^{\lambda}\,^{(4)}\!K^{\tau}_{\;\lambda\tau}-
^{(4)}\!\!K_{\tau\lambda}^{\;\;\;\lambda}\,^{(4)}\!K^{\tau\lambda}_{\;\;\;\lambda}+
^{(4)}\!\!K_{\tau\gamma\lambda}\,^{(4)}\!K^{\tau\lambda\gamma}
\Big)=0,\label{2111} \ea where $V$ denotes the ``volume'' of the
internal space. Note that for $\xi=0$, the torsion terms do not
influence the general sum rules in the present case. In fact, for
$\xi=0$ it follows that \ba
\frac{V\Lambda}{2\kappa_{5}^{2}}+\lambda+\tilde{\lambda}=0,\label{22}
\ea which states that, for non-warped branes, it is possible to
exist an AdS$_{5}$ bulk, even for strictly positive tension values
associated with the branes. Another interesting case is obtained
for $\xi=-1$. In such case the bulk cosmological constant is
factored out and consequently \ba
\Big(\,{}^{(4)}\!\!K_{\tau\lambda}^{\;\;\;\lambda}\,^{(4)}\!K^{\tau\lambda}_{\;\;\;\lambda}-
^{(4)}\!\!K_{\tau\gamma\lambda}\,^{(4)}\!K^{\tau\lambda\gamma}-2D^{\lambda}\,^{(4)}\!K^{\tau}_{\;\lambda\tau}
\Big)=\frac{4\kappa_{5}^{2}}{V}(\lambda+\tilde{\lambda}).\label{23}
\ea Note that the left hand side (LHS) of (\ref{23}) can be
interpreted as the difference between
$\mathring{\bar{R}}$ and $\bar{R}$. In other words, the LHS of
Eq.(\ref{23}) measures the contribution of the torsion terms to
the brane curvature, i. e., it indicates how much the brane
curvature differs itself from zero, due to torsion terms. So, we
can write schematically \ba
\mathring{\bar{R}}-\bar{R}=\frac{4\kappa_{5}^{2}}{V}(\lambda+\tilde{\lambda}).\label{24}
\ea We see that the effect of the torsion in the brane curvature
is proportional to the branes tension values in the two branes
scenario, but it decreases with the distance between the branes.
Moreover, since $\kappa_{5}^{2}=8\pi G_{5}\sim 1/M_{5}^{3}$, such
an effect is about $1/(VM_{5}^{3}$). Therefore, it indicates the
low magnitude of torsion effects in the braneworld scenario with
large extra transverse dimension, since it is suppressed by the
5-dimensional Planck scale and also by the volume of the internal
space. Obviously, in a braneworld scenario which solves the
hierarchy problem the typical scale of the higher dimensional
Planck mass is of order $M_{5}\sim M_{weak}$ and then, the
suppression due to the internal space volume is attenuated.

\subsection{The warped case}

In the absence of a factorizable geometry, some configurations of the
warp factor may be responsible for the right mass partition in the
Higgs mechanism without the necessity of any additional hierarchy
\cite{RS}. Starting from the general Eq.(\ref{19}), we shall look
at the most important cases, namely $\xi=-1,0,1$.

For $\xi=-1$ we have \ba
\Big(\,^{(4)}\!\!K_{\tau\lambda}^{\;\;\;\lambda}\,^{(4)}\!K^{\tau\lambda}_{\;\;\;\lambda}-
^{(4)}\!\!K_{\tau\gamma\lambda}\,^{(4)}\!K^{\tau\lambda\gamma}-2D^{\lambda}\,^{(4)}\!K^{\tau}_{\;\lambda\tau}
\Big)=\frac{4\kappa_{5}^{2}}{\oint
W^{-2}}(\lambda+\tilde{\lambda}).\label{25} \ea This is the warped
analogue of Eq.(\ref{23}) with the volume of the internal space
replaced by the circular integral of $W^{-2}$ in the denominator
of the right hand side. The same conclusions as the $\xi=-1$ case
of the previous Subsection still hold, but here we call the
attention to the minuteness of the torsion terms: even
contributing with such low magnitude effect to the brane
curvature, it allows the branes to have both the same sign
associated to their respective tension values.

The bulk spacetime type can be better visualized in the $\xi=0$
case. Since all torsion terms of Eq.(\ref{19}) are factored out,
it follows that \ba \frac{\Lambda}{2\kappa_{5}^{2}}\oint W+\lambda
W_{\lambda}+\tilde{\lambda}W_{\tilde{\lambda}}=0.\label{26}\ea
Therefore, as $\oint W<0$, it is easy to see that if $\lambda
,\tilde{\lambda}>0$ then necessarily $\Lambda>0$ corresponding to
an dS$_{5}$ bulk geometry. Otherwise, being $\lambda
,\tilde{\lambda}<0$ one arrives at an AdS$_{5}$ bulk geometry.

For the $\xi=1$ case, a slight modification of Eq.(\ref{25})
deserves a notification. The implementation of $\xi=1$ in the
Eq.(\ref{19}) results in \ba \frac{\Lambda}{\kappa_{5}^{2}}\oint
W^{2}+\lambda
W_{\lambda}^{2}+\tilde{\lambda}W_{\tilde{\lambda}}^{2}-\frac{V}{4\kappa_{5}^{2}}
\Big(2D^{\lambda}\,^{(4)}\!K^{\tau}_{\;\lambda\tau}-
^{(4)}\!\!K_{\tau\lambda}^{\;\;\;\lambda}\,^{(4)}\!K^{\tau\lambda}_{\;\;\;\lambda}+
^{(4)}\!\!K_{\tau\gamma\lambda}\,^{(4)}\!K^{\tau\lambda\gamma}\Big)
=0.\label{27} \ea Now, isolating the torsion contribution to the
curvature we have \ba
\Big(2D^{\lambda}\,^{(4)}\!K^{\tau}_{\;\lambda\tau}-
^{(4)}\!\!K_{\tau\lambda}^{\;\;\;\lambda}\,^{(4)}\!K^{\tau\lambda}_{\;\;\;\lambda}+
^{(4)}\!\!K_{\tau\gamma\lambda}\,^{(4)}\!K^{\tau\lambda\gamma}\Big)=\frac{4\Lambda}{V}\oint
W^{2}+\frac{4\kappa_{5}^{2}}{V}(\lambda
W_{\lambda}^{2}+\tilde{\lambda}W_{\tilde{\lambda}}^{2}).\label{28}\ea
From  Eq.(\ref{28}) we see that the torsion contribution to the
brane curvature is constrained by the internal space volume,
however terms coming from the warped compactification --- as
$\oint W^{2}$ and $\oint W_{\lambda,\tilde{\lambda}}^{2}$ --- can turn this
contribution more appreciable. In particular, the first term of
the right hand side of (\ref{28}) is the dominant one, since it is
not suppressed by the 5-dimensional Planck scale and it is
multiplied by the bulk cosmological constant. We shall make more
comments about these results in the next Section.

\section{Concluding remarks and outlook}

There are some alternative derivations of the junction conditions
for a brane in a 5-dimensional bulk, when Gauss-Bonnet equations
are used to describe gravity \cite{deruele}. Also, Israel junction
conditions can be generalized for a wider class of theories by
direct integration of the field equations, where a specific
non-minimal coupling of matter to gravity suggests promising
classes of braneworld scenarios \cite{japa}.  In addition, it is
also possible to generalize matching conditions for cosmological
perturbations in a teleparallel Friedmann universe, following the
same lines as \cite{deru}.

In the case studied here, however, the matching conditions are not
modified by the inclusion of torsion terms in the connection. As
noted, it is a remarkable and unexpected characteristic. Besides,
all the development concerning the formalism presented is
accomplished in the context of braneworld models. In such
framework, the appearance of torsion terms is quite justifiable.
However, the fact that the matching conditions remain unalterable
in the presence of torsion is still valid in usual 4-dimensional
theories.

Once investigated the junction conditions, we have obtained, via
Gauss-Codazzi formalism, the Einstein effective projected equation
on the brane. If, on one hand, the torsion terms do not intervenes
in the usual Israel-Darmois conditions, on the other hand it
modifies drastically the brane Einstein equations. Eq.(\ref{c10})
shows up the strong dependence of the new effective cosmological
constant on the four and five-dimensional contorsion terms. It
reveals promising possibilities. For instance, by a suitable
behavior of such new terms, $\tilde{\Lambda}_{4}$ can be very
small. In a more complete scenario, $\tilde{\Lambda}_{4}$ could be
not even a constant. It must be stressed that these types of
modification in the projected Einstein equation also appear in
other models in modified gravity \cite{NDN}.

This Chapter intends to give the necessary step in order to formalize
the mathematical implementation of torsion terms in braneworld
scenarios. The application of our results are beyond the scope of
this work. We finalize, however, pointing out some interesting research
lines coming from the use of the results --- obtained in this work --- in cosmological
problems.

The final result is very important from the cosmological viewpoint. It is clear that deviations of the usual
braneworld cosmology can be obtained from the analysis of phenomenological systems
in the light of Eq.(\ref{c9}). Physical aspects, more specifically
the analysis of cosmological signatures as found in ref.
\cite{OC}, arising from the combination of the extra dimensions and
torsion should be systematically investigated and compared with
usual braneworld models. The ubiquitous presence of torsion terms
leads, by all means, to subtle but important deviations of usual
braneworlds in General Relativity. For instance, the equation
(\ref{c9}) can be used as a starting point to describe the flat
behavior of galaxy rotational curves without claim for dark
matter. This last problem was already analyzed in the context of
brane worlds \cite{ROTGAL}, however the outcome arising from the
torsion terms has never been investigated. A systematic
comparative study between usual braneworld models and those braneworld models
embedded in an Einstein-Cartan manifold is, potentially,
interesting since it can lead us to new branches inside brane
physics. We shall address to those questions in the future.

Two additional remarks must be pointed out. First, all
the development concerning the formalism presented is accomplished in the context of braneworld
models. In such framework, the appearance of torsion terms is quite
justifiable. However, the fact that the matching conditions remain
unalterable in the presence of torsion is still valid in usual 4-dimensional theories. Second,
the discontinuity orthogonal to the brane is analyzed, since it is the
unique possibility: the brackets $[A]$ of any quantity $A$
denote, by definition, the jump across the brane. The
geometric reason points in the same direction; the connection must
be continuous along the brane in order to guarantee the full
applicability of standard calculations on the brane that works as a model to the universe.

This work concerns some effects evinced by torsion terms
corrections --- both in the bulk and on the brane. To study a
typical gravitational signature arising from a gravitational
system we performed in Section II the analysis based upon the well
known Taylor expansion tools --- strongly reminiscent of the
assumption of a direction orthogonal to the brane --- of the bulk
metric in terms of the brane metric, taking into account bulk
torsion terms. Our main result is summarized by Eq.(\ref{modif}).
It shows how the bulk torsion terms intervene in the black hole
area, in an attempt to find some observable effects arising from
the torsion properties. Again, its highly non-trivial form can be
better studied in the context of a specific model. It is out of
the scope of this paper, nevertheless we shall point a line of
research in this area. It could be interesting to apply the
results found in this Section to some gravitational systems, in
analogy to what was accomplished in standard braneworld scenarios
(see, for instance, references \cite{merc}).

In order to study the behavior of the brane torsion terms we
extend, in Section IV, the braneworld sum rules. It was
demonstrated that the consistency conditions do not preclude the
possibility of torsion on the brane. It was shown, however, that
the torsion effects in the brane curvature are suppressed. It
could, in principle, explain a negative result for experiments
from the geometrical point of view. Just for a comparative
complement, in reference \cite{SEN} the 5-dimensional torsion
field, identified with the rank-2 Kalb-Ramond (KR) field, was
considered in the bulk. It was demonstrated by the authors the
existence of an additional exponential damping for the zero-mode
of the KR field arisen from the compactification of the transverse
dimension. In some sense, our purely geometrical sum rules
complete the analysis concerning the presence of torsion, this
time on the brane.

In this paragraph we would like to call attention for some related
issues appearing in the literature. In \cite{SEN} it was
shown that in an effective 4-dimensional theory on the visible
brane, the KR field --- as a source of torsion --- is suppressed
when a torsion-dilaton-gravity action in a Randall-Sundrum
braneworld scenario is considered, explaining the apparent
insensitivity of torsion in the brane. It was shown, however, that
even in this case the KR field may led to new signatures in TeV
scale experiments, when a coupling between dilaton and torsion is
taken into account. The warped extra-dimensional formalism points
to the presence of new interactions, of significant
phenomenological importance, between the Kaluza-Klein modes of the
dilaton and the KR field.

Briefly speaking, the results of this paper point to the fact that
the hypothesis of a torsionless brane universe may be based upon a
justified impression, since its effects from the bulk (studied
from a quantum field theory approach) and from the brane (analyzed
via the geometrical sum rules) are suppressed by some damping
factor. We emphasize, however, that in the context above, the
naive estimative of the 4-dimensional torsion effects (\ref{28})
must be complemented by the results of a more specific system via Eq. (\ref{modif}).
Such a characterization may put this gravitational and geometrical
approach in the same level, concerning the brane torsion
phenomenology, as, for instance, the massive spectrum of
5-dimensional KR field signature which can be viewed in a
TeV-scale accelerator \cite{SEN}. In this vein, the torsionless
brane universe may be naturally substituted by a more fidedigne
braneworld scenario that contains torsion, and may be useful to a
more precise description of physical theories.

\section*{Acknowledgment}
R. da Rocha thanks CNPq 304862/2009-6 for financial support.

\end{document}